%% file: main.tex
\documentclass[%
 aip,
 amsmath,amssymb,
 reprint,%
]{revtex4-1}

\usepackage{graphicx}
\usepackage{dcolumn}
\usepackage{bm}
\usepackage{xcolor}  
\usepackage[utf8]{inputenc}
\usepackage[T1]{fontenc}
\usepackage{mathptmx}
\usepackage{etoolbox}

\makeatletter
\def\@email#1#2{%
 \endgroup
 \patchcmd{\titleblock@produce}
  {\frontmatter@RRAPformat}
  {\frontmatter@RRAPformat{\produce@RRAP{*#1\href{mailto:#2}{#2}}}\frontmatter@RRAPformat}
  {}{}
}%
\makeatother
\begin{document}

\preprint{AIP/123-QED}

\title[Three-core fiber Fabry-Perot resonator for dual-frequency comb generation]{Three-core fiber Fabry-Perot resonator for dual-frequency comb generation}
\author{Thomas Bunel}
 \email{thomas.bunel@ulb.be}
\affiliation{University of Lille, CNRS, UMR 8523-PhLAM Physique des Lasers, Atomes et Molécules, F-59000, Lille, France 
}%
\affiliation{Service OPERA-Photonique, Université libre de Bruxelles (ULB), Brussels, Belgium}

\author{Antonio Cutrona}
\affiliation{Emergent Photonics Research Centre, Dept. of Physics, Loughborough University, Loughborough, LE11 3TU, England, UK}%

\author{Debanuj Chatterjee}
\author{Damien Labat}
\author{Vincent Andrieux}
\author{Geraud Bouwmans}
\author{Andy Cassez}
\affiliation{University of Lille, CNRS, UMR 8523-PhLAM Physique des Lasers, Atomes et Molécules, F-59000, Lille, France 
}%

\author{Antonin Moreau}
\author{Julien Lumeau}
\affiliation{Aix Marseille Univ., CNRS, Centrale Med, Institut Fresnel, Marseille, France
}%

\author{Manal Arbati}
\author{Alexis Bougaud}
\author{Benjamin Wetzel}
\affiliation{XLIM Research Institute, CNRS UMR 7252, Université de Limoges, Limoges, France}%

\author{Alessia Pasquazi}
\affiliation{Emergent Photonics Research Centre, Dept. of Physics, Loughborough University, Loughborough, LE11 3TU, England, UK}%

\author{Matteo Conforti}
\author{Arnaud Mussot}
\affiliation{%
University of Lille, CNRS, UMR 8523-PhLAM Physique des Lasers, Atomes et Molécules, F-59000, Lille, France 
}%
 
\date{\today}

\begin{abstract}
Fiber Fabry-Perot resonators have proven their ability to generate broad and stable optical frequency combs, and are ideal devices for fiber systems as they are high-Q, compact, and easily integrated with FC/PC connectors. Here, we present an advanced fiber Fabry-Perot resonator designed for multi-frequency comb generation and spatial multiplexing. The resonator is fabricated using a three-core optical fiber and is able to generate two mutually coherent frequency combs while being locked to a driving laser. Multiplexing of the combs is achieved with a fan-in/fan-out system, enabling a fully fiber-based experimental setup. The generated combs, induced by cavity solitons, feature a 1.27~GHz repetition rate and a bandwidth above 40~nm. A slight difference in the group index of each core leads to a 112~kHz repetition rate offset between the combs, enabling dual-comb spectroscopy proof-of-concept measurement of a 0.1~nm absorption band.
\end{abstract}

\maketitle

%

\section{Introduction}

Since their demonstration in the late 20\textsuperscript{th} century, optical frequency combs (OFCs) have revolutionized precision measurements of time and frequency~\cite{jones_carrier-envelope_2000,holzwarth_optical_2000,diddams_optical_2001}. Remarkable applications have been enabled by dual-frequency-comb which leverage the coherence properties of OFC for rapid broadband spectral analysis with high accuracy. This approach has transformed fields such as spectroscopy~\cite{coddington_dual-comb_2016,picque_frequency_2019}, lidar~\cite{coddington_rapid_2009,suh_soliton_2018,xu_review_2025}, and hyperspectral imaging~\cite{vicentini_dual-comb_2021,martin-mateos_direct_2020}. When two OFCs with slightly different repetition rates interfere on a photodiode, a radio-frequency (RF) comb is generated, consisting of heterodyne beats between pairs of optical comb teeth. This RF comb is readily accessible with RF electronics and contains the relevant spectral or temporal information of the optical combs~\cite{coddington_dual-comb_2016,picque_frequency_2019,coddington_rapid_2009,suh_soliton_2018}. Various technological strategies have been developed to generate two mutually coherent OFCs, a key requirement for dual-comb applications. These include phase-locking of two mode-separated lasers~\cite{link_dual-comb_2015}, bidirectional lasers~\cite{ideguchi_kerr-lens_2016,Li:20,mehravar_real-time_2016}, arrays of electro-optic modulators (EOMs) driven by a common laser~\cite{bancel_all-fiber_2023,parriaux_electro-optic_2020}, and multi-frequency comb generation in Kerr resonators~\cite{suh_soliton_2018,zhang_soliton_2023,lucas_spatial_2018,xu_dual-microcomb_2021,bunel_dual-frequency_2025}. Among these, microresonators have attracted particular attention over the past decade~\cite{sun_applications_2023,pasquazi_micro-combs_2018}, owing to their ability to be integrated on a chip and capability to generate high-repetition-rate OFCs. Most importantly, their coherent driving schemes directly stabilize the OFC phase~\cite{lei_optical_2022}, eliminating the need for complex mode-locking mechanisms.
This feature has enabled the generation of mutually coherent frequency combs when pumped by the same driving laser~\cite{bao_microresonator_2019,dutt_-chip_2018,suh_microresonator_2016,trocha_ultrafast_2018,Rebolledo_2023}, with coherence further enhanced when both OFCs are generated within a single monolithic resonator, thereby sharing the same noise source. Such systems have been realized in different configurations, including counter-propagating waves~\cite{suh_soliton_2018,zhang_soliton_2023}, orthogonal polarizations~\cite{xu_dual-microcomb_2021}, and distinct transverse modes~\cite{lucas_spatial_2018}. In addition to chip-integrated micro-ring resonators, new fiber-based Fabry–Perot resonators have emerged as a promising alternative. They offer easy integration into fiber systems through low-loss FC/PC connectors, compact footprints, high Q-factors reaching up to hundreds of millions~\cite{bourcier_optimization_2024,bourcier_investigation_2025} and the ability to generate Kerr frequency combs with GHz repetition rates~\cite{obrzud_temporal_2017,nie_synthesized_2022,li_ultrashort_2023,bunel_28_2024,bunel_brillouin-induced_2025}. These systems have also demonstrated potential for imaging~\cite{xu_toward_2025} and dual-comb generation whether with two independent cavities in self-injection locking~\cite{Qin:25}, or with a monolithic architectures through pulsed pumping configuration~\cite{bunel_dual-frequency_2025}. Their simple design, consisting of an optical fiber terminated by two ferrules with dielectric mirrors deposited on each facet, makes fabrication highly versatile, given the wide range of available fiber types. At the same time, multicore fibers, initially developed for telecommunications~\cite{morioka_enhancing_2012}, have emerged as an excellent platform for generating independent combs with high mutual coherence, as demonstrated through electro-optic OFC spectral broadening in each core during single-pass propagation~\cite{bancel_all-fiber_2023,chatterjee_sensitivity_2025,Chatterjee2025}.

\begin{figure*}
\includegraphics[width=\linewidth]{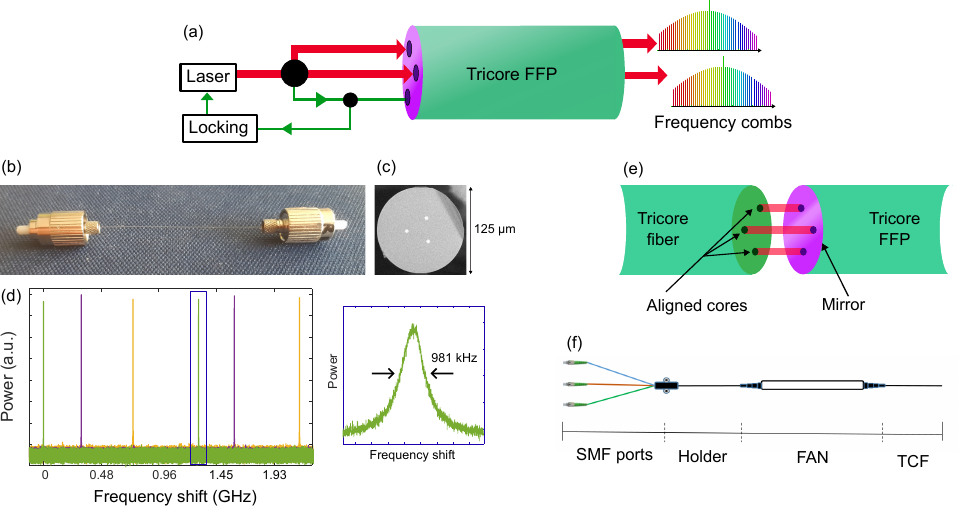}
\caption{Principle of dual-Kerr frequency comb generation in a three-core fiber Fabry-Perot resonator. (a)~Schematic representation of cavity locking and combs generation. (b)~Three-core FFP resonator used in this work. (c)~SEM view of a section of the TCF. (d)~Schematic view of the cavity coupling and cores alignment. (e)~Transfer function of each resonator core; purple line: Core~1; orange line: Core~2; green line: Core~3; blue inset: zoom-in of Core~3 resonance. (f)~Schematic of a fan.}
\label{cavite}
\end{figure*}

In this work, we present an advanced multicore fiber Kerr resonator for multi-frequency comb generation from a single continuous wave (CW) pump. A dual-comb is generated by exploiting the multicore architecture and fine control of the core separation of a single fiber Fabry-Perot (FFP) resonator. 
In an all-fiber experimental setup, the FFP resonator fabricated from a three-core fiber is used to simultaneously generate two mutually coherent OFCs induced by cavity solitons (CS) in two different cores, while the third one is employed to lock the pump laser to the cavity. As a result, the two OFCs remain mutually coherent without interacting with one another or disturbing the stabilization signal. Finally, we demonstrate the potential of this dual-comb source through a proof-of-concept spectroscopy experiment, measuring the transfer function of a programmable optical filter configured as a notch filter.

\section{Results}\label{RESULTS}

The use of a three-core resonator for dual-Kerr frequency comb generation is highly advantageous, as each core can be regarded as an independent cavity while still being part of a common system, and therefore subject to the same perturbations. 
Due to their very narrow resonance linewidths in the kHz range, frequency combs generated in fiber cavities are highly sensitive to even the slightest variations, whether originating from the cavity itself because of vibrations or temperature fluctuations, or from variations in the pump laser. Cavity locking is therefore required to maintain a stable cavity detuning between the pump and the cavity resonance, a key parameter for frequency comb generation~\cite{englebert_high_2023,bunel_28_2024}.
Our approach is summarized in Fig.~\ref{cavite}(a), where a single CW laser and a single three-core FFP resonator are employed: the laser is split into three branches, two for pumping and OFC generation in two separate cores, and one for cavity locking in the third core. This configuration ensures the long-term generation of two mutually coherent OFCs, stabilized through the third cavity core.

\subsection{The three-core fiber Fabry-Perot resonator}\label{CAVITY}

The FFP resonator [Fig.~\ref{cavite}(b)] is fabricated from an 8.07~cm three-core fiber (TCF). This fiber consists of a silica cladding with three silica cores arranged in a triangular configuration, separated by 30~\textmu{}m [Fig.~\ref{cavite}(c)]. This spacing ensures that propagation remains as similar as possible across the three channels, while still providing sufficient separation to limit inter-core crosstalk~\cite{bancel_all-fiber_2023}. The cores exhibit nearly identical anomalous group velocity dispersion (GVD) and nonlinear coefficient, with $\beta_2=-10$~ps$^2$km$^{-1}$ and $\gamma=5$~W$^{-1}$km$^{-1}$, respectively. The total fiber diameter (core + cladding) is 125~\textmu{}m, allowing compatibility with standard FC/PC connectors where dielectric mirrors are deposited. The transfer function of each core [Fig.~\ref{cavite}(d)] is measured using an optical phase modulator for which driving frequency is swept over more than one free spectral range (FSR) in the same way as in~\cite{bunel_impact_2023}. Each core exhibits a very high and almost similar finesse (or Q-factor) with sub-MHz resonance linewidth [inset in Fig.~\ref{cavite}(d)], and the FSR are about 1.275~GHz with slight variations. These differences are summarized in Table~\ref{tab:cores} and can be attributed to inhomogeneities during the fabrication process of the cavity, due to both mirror imperfection and slight group index variations between the cores.

\begin{table}
    \centering
    \caption{Cavity cores characteristics}
    \begin{tabular}{|c|c|c|c|}
    \hline
        &Core~1&Core~2&Core~3\\
        \hline
        $FSR$&1.2755~GHz&1.2757~GHz&1.2758~GHz\\
        $\mathcal{F}$&1300&1180&980\\
        $Q$&$197\cdot10^6$&$179\cdot10^6$&$148\cdot10^6$\\
        \hline
    \end{tabular}
    \label{tab:cores}
\end{table}

Independent assignment of light to each core and demultiplexing at the FFP resonator output [Fig.~\ref{cavite}(e)] are achieved using the fans shown in Fig.~\ref{cavite}(f). The fans (manufactured by \textit{Chiralphotonics}) consist of three single-mode fibers (SMF-28) that converge into a TCF, which is then butt-coupled to the resonator. The insertion loss is 2~dB per fan. The FC/PC connectors of both the resonator and the TCF fan are joined using mating sleeves, after which a controlled rotation is applied to ensure proper core alignment between the input (or output) TCF and the FFP resonator [Fig.~\ref{cavite}(e)], with less than 4~dB total injection loss for all cores.

\subsection{Optical frequency comb generation}\label{OFC}

\begin{figure}
\includegraphics[width=8.5cm]{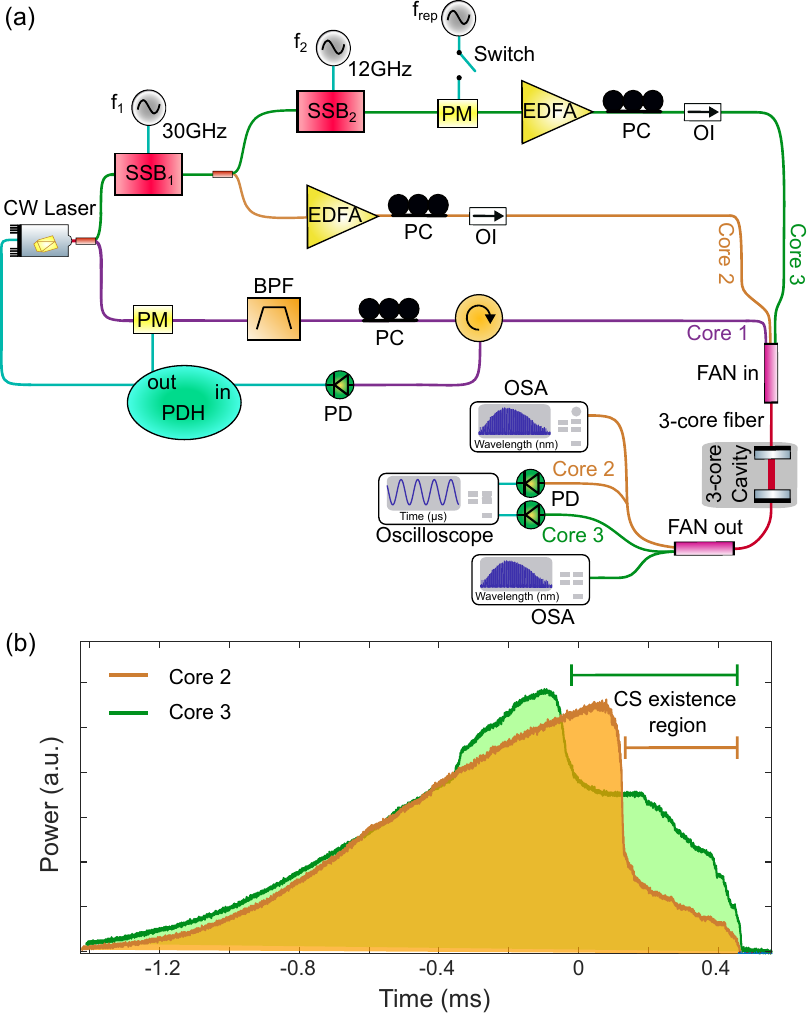}
\caption{Experimental setup. (a)~Scheme of the all-fiber setup. Phase modulation in Core~3 facilitates single-soliton triggering and is switched off once the soliton is formed. CW: Continuous Wave; SSB: Single-Sideband generator; PM: Phase Modulator; EDFA: Erbium Doped Fiber Amplifier; PC: Polarization Controller; OI: Optical Isolator; BPF: Bandpass Filter; PD: Photodiode (1~GHz bandwidth) ; PDH: Pound-Drever-Hall; OSA: Optical Spectrum Analyzer. (b)~Nonlinear transfer function in Cores~2 and 3.}
\label{setup}
\end{figure}

The \{fan-in/FFP resonator/fan-out\} system is integrated into the all-fiber experimental setup shown in Fig.~\ref{setup}(a). A CW laser is split into three branches: a control beam (purple line), which locks the laser frequency to a cavity resonance of Core~1 using a Pound–Drever–Hall (PDH) system~\cite{drever_laser_1983,black_introduction_2001}, and two pump beams (green and orange lines) that generate OFCs in Cores~2 and 3, respectively. The pump beams are frequency-shifted using single-sideband generators [SSB$_1$ and SSB$_2$ in Fig.~\ref{setup}(a)] according to frequency synthesizer values ($f_{1}$ and $f_{2}$). This allows to achieve the correct cavity detuning between the pump and the cavity resonance, enabling independent OFC generation in Cores~2 and 3. Erbium-doped fiber amplifiers (EDFA) boost the pump powers to 400~mW and 600~mW in Cores~2 and 3, respectively. This difference in pump power was chosen experimentally and is consistent with the difference in finesse [Table~\ref{tab:cores}]. Polarization controllers (PC) and optical isolators (OI) are included in each pump arm to align the polarization with the corresponding cores and suppress back reflections, respectively. After demultiplexing with the fan-out, the generated signals are characterized in both the time and spectral domains using optical spectrum analyzer (OSA), electrical spectrum analyzer (ESA), and oscilloscope.
Note that the SSBs are intentionally not placed symmetrically in the two pump arms [green and orange lines in Fig.~\ref{setup}(a)]. Specifically, SSB$_1$ shifts both pump beams relative to the control beam (purple line), while SSB$_2$ shifts the Core~3 pump beam (green line) relative to the Core~2 pump beam (orange line). This configuration ensures that both pump beams experience frequency variations similar to those of the control arm and enables long-term cavity soliton generation in both cores using the same detuning scan-and-stop method~\cite{herr_temporal_2014}. Once the laser and cavity are locked via PDH, resonances in Cores~2 and 3 are scanned by sweeping the frequency of SSB$_1$ over more than one FSR ($f_{1}$ sweeps from 29 to 31~GHz). Then, $f_{2}$ is tuned so that the resonances of Cores~2 and 3 coincide in time during the $f_{1}$ frequency sweep [Fig.~\ref{setup}(b)]. These scans exhibit characteristic steps associated with cavity soliton formation, as indicated in Fig.~\ref{setup}(b)~\cite{pasquazi_micro-combs_2018,herr_temporal_2014}. Care is taken to temporally align these soliton steps during the scan, ensuring that for specific values of $f_{1}$, cavity solitons are generated simultaneously in Cores~2 and 3. Thereby, two OFCs are generated at the same time with a common scan-and-stop induced by SSB$_1$. Finally, a phase modulator (PM), driven at a frequency matching the core’s FSR with a modulation amplitude of 0.18~rad, is inserted in the Core~3 pump arm [green line in Fig.~\ref{setup}(a)] to facilitate single-soliton triggering~\cite{englebert_high_2023,jang_temporal_2015}. This phase modulation is required to suppress stimulated Brillouin scattering (SBS), which strongly resonates in this core during the cavity scan for soliton excitation. Once a single soliton is established, the PM is switched off to maintain CW pumping. Interestingly, SBS does not appear to have a significant impact in Core~2, where single-soliton generation occurs much more easily without requiring phase modulation. The Brillouin effect plays a key role in the nonlinear dynamics of FFP cavities~\cite{jia_photonic_2020,nie_synthesized_2022,bunel_brillouin-induced_2025}, and its influence strongly depends on the resonance position relative to the Brillouin gain. A detailed investigation of SBS impact on cavity solitons will be presented in future works.

\begin{figure*}
\includegraphics[width=\linewidth]{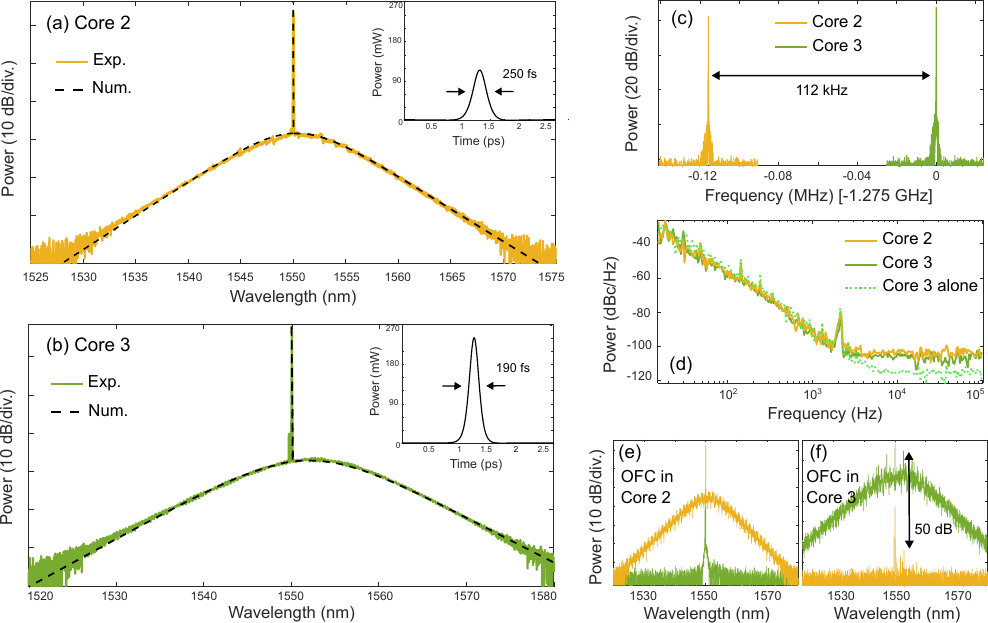}
\caption{Optical frequency combs generation. (a) and (b)~OFC generation induced by single cavity soliton in Core 2 and 3, respectively, in good agreement with numerics. The insets are representations of the time domain solitons corresponding to numerics in black dotted lines. 
(c)~RF beatnotes of OFC generated in Core~2 and 3, centered at the repetition rate of Core~3 OFC. (d)~Phase noise spectra measurements of OFC generated in Core 2 and 3, corresponding to signals shown in (b) and (c). (e) and (f)~Crosstalk measurement between Core 2 and 3.}
\label{combs}
\end{figure*}

By setting $f_{2}=11.1293$~GHz and scanning $SSB_1$ to reach $f_{1}=30.9644$~GHz, corresponding to detunings of 0.028~rad in Core~2 and 0.065~rad in Core~3, single-CS frequency combs are simultaneously generated in both cores, as shown in Fig.~\ref{combs}(a) and (b). The resulting OFCs span over 40~nm in Core~2 and 60~nm in Core~3.
Both measurements are in excellent agreement with numerical simulations [black dashed lines] based on the Lugiato–Lefever equation adapted for Fabry–Perot resonators~\cite{cole_theory_2018,ziani_theory_2023}, including the Raman response~\cite{bunel_experimental_2025}. These numerics yield the corresponding time traces shown in the insets of Fig.~\ref{combs}(a) and (b), with pulse durations of 250~fs and 190~fs (full width at half maximum), respectively. For a more detailed characterization of the generated combs, we measured the beatnote of each OFC with an ESA [Fig.~\ref{combs}(c)]. We find a 112~kHz difference in the repetition rates, consistent with the FSR difference [see Table~\ref{tab:cores}]. For Core~3, the spectral recoil observed in Fig.~\ref{combs}(b) is characteristic of a Raman frequency shift~\cite{karpov_raman_2016}, which modifies the repetition rate by causing the soliton to drift relative to the intracavity field~\cite{suh_soliton_2018}. However, based on the spectrum in Fig.~\ref{combs}(b), the numerically calculated Raman-induced change in repetition rate is only 2~kHz (not shown). Compared to the 112~kHz measured, the Raman effect alone thus cannot account for the frequency difference observed in the beatnotes of Fig.~\ref{combs}(c) as observed in other dual-Kerr comb systems~\cite{suh_soliton_2018}; it is primarily attributed to variations in the FSR arising from imperfections in the resonator fabrication, \textit{i.e.}, effective index differences of $10^{-4}$ order.

\begin{figure*}
\includegraphics[width=\linewidth]{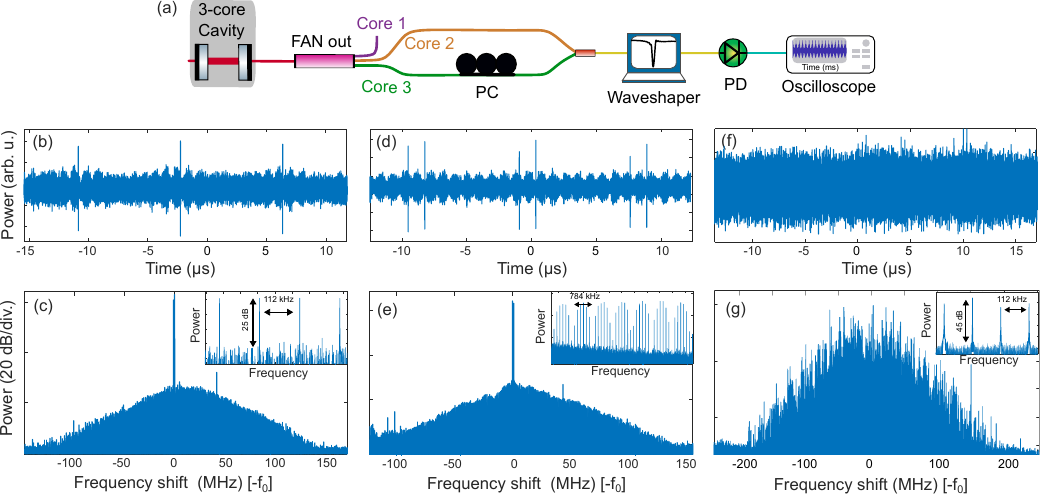}
\caption{Dual-frequency comb generation. (a)~Cavity output dual-comb setup. (b), (d) and (f)~Interferograms in the time domain. (c), (e) and (g)~Fourier transforms of (b), (d) and (f), respectively; the insets are zoom-ins to show the comb structures. (b)-(c): single soliton in each core. (d)-(e): single-soliton in Core~2 and soliton-state composed of two solitons in Core~3. (f)-(g): multi-solitons regime in each core.}
\label{dc}
\end{figure*}

Phase-noise spectra of the repetition rates for each OFC are reported in Fig.~\ref{combs}(d). Measurements were performed both when a single OFC was generated and when the two OFCs were generated simultaneously. In the dual-comb configuration, both combs exhibit similar phase noise and excellent stability [orange and green lines in Fig.~\ref{combs}(d)], with phase-noise levels of $-100$~dBc/Hz at 1~kHz and $-40$~dBc/Hz at 10~Hz. The only noticeable difference is a slightly higher noise floor in Core~2. Also, as previously observed in another FFP resonator dual-comb scheme~\cite{bunel_dual-frequency_2025}, the background phase noise increases when both OFCs are generated simultaneously, reflecting an increase in white noise during the multi-frequency comb generation process. As shown by the green dashed line in Fig.~\ref{combs}(d), the phase noise of the OFC generated in Core~3 is 10~dB lower when Core~2 is off, for frequencies above 10~kHz. Similar results were obtained for Core~2 (not shown here).

Finally, the crosstalk of the entire system \{fan-in/FFP resonator/fan-out\} is measured to be below 30~dB, as shown in Fig.~\ref{combs}(e) and (f), where the outputs of both cores are recorded while only one OFC is generated, in Core~2 in Fig.~\ref{combs}(e) and Core~3 in Fig.~\ref{combs}(f). The low crosstalk ensures that each core can be operated independently, without the field propagating in one core affecting those in the others. This greatly facilitates OFC generation and stabilization, even though all cores are integrated within the same 125~\textmu{}m diameter fiber.

\subsection{Dual-frequency comb experiment}\label{DC}

The difference in repetition rate between the two generated combs is required in dual-comb systems as it enables the down-conversion process~\cite{coddington_dual-comb_2016,picque_frequency_2019,coddington_rapid_2009}. The interferogram between the two OFCs generated in Cores~2 and 3 [shown in Fig.~\ref{combs}(a) and (b)] is obtained by mixing the combs using a 50/50 coupler [Fig.~\ref{dc}(a), without the \textit{Waveshaper}] and is shown in Fig.~\ref{dc}(b). Its Fourier transform is displayed in Fig.~\ref{dc}(c), revealing a characteristic $\mathrm{sech}^2$ profile corresponding to the product of the two optical spectra and a magnification factor of $m = f_{rep} / \Delta f = 1.134 \times 10^{4}$ (where $f_{rep} = 1.275$~GHz is the repetition rate and $\Delta f = 112$~kHz is the repetition rate difference). A clear comb structure highlights the high mutual coherence between the two combs [inset in Fig.~\ref{dc}(c)], with a tooth spacing equal to $\Delta f$. A pretty good signal-to-noise ratio exceeding 25~dB near the RF carrier further confirms that this dual-comb source is a good candidate for metrological applications. The carrier frequency offset between the two combs is set by the SSB$_2$ frequency $f_2$, which determines the RF comb center frequency $f_0$. In this case, $f_0$ = 352~MHz, in perfect agreement with the expected relation $f_0 = p \times f_{\mathrm{rep}} - f_{2}$, with $p \in \mathbb{N}$ (\textit{i.e.}, $352~\mathrm{MHz} = 9 \times 1.2757~\mathrm{GHz} - 11.1293~\mathrm{GHz}$)~\cite{xu_dual-microcomb_2021,bunel_dual-frequency_2025}, and within the photodiode bandwidth.
Thanks to the multiplexing setup, it is possible to independently control the nonlinear operating regime of each core. For instance, Fig.~\ref{dc}(d) and (e) show the case where two solitons are generated in Core~2 while a single-soliton is generated in Core~3. In this case, the pulses from Core~2 overlap twice with those from Core~3 over a period of $1/\Delta f$ [Fig.~\ref{dc}(d)], enabling the detection of short spectral modulations of 784~kHz in the RF domain [Fig.~\ref{dc}(e)], \textit{i.e.} $m \times 784~\mathrm{MHz}=8.9~\mathrm{GHz}$ in the optical domain. Note that this 8.9~GHz spectral modulation corresponds to a $1/8.9~\mathrm{GHz}=$112~ps, \textit{i.e.} 2.31~cm soliton separation in Core~3. Therefore, in addition to its potential for metrology, this device is also suitable for self-imaging of its nonlinear dynamics and real-time cavity soliton visualization~\cite{yi_vahala_2018}. Finally, Fig.~\ref{dc}(f) and (g) illustrate the case of multi-soliton states in both cores. Here, the interferogram no longer consists of sharp pulses but rather of a soliton gas [Fig.~\ref{dc}(f)]. Nevertheless, owing to the periodicity and the stability of the soliton gas, the RF spectrum obtained from the Fourier transform remains a comb with teeth separated by the repetition rate difference, $\Delta f = 112$~kHz [inset in Fig.~\ref{dc}(g)]. Although the RF comb appears scrambled due to the multi-soliton interferences, it remains stable over time and exhibits a significantly higher signal-to-noise ratio, 20~dB higher compared to the single-CS regime [insets of Fig.~\ref{dc}(c) and (g)], owing to the higher optical spectral power available in multi-soliton regimes. The stability of these soliton gases demonstrates, on the one hand, that solitons within the same core do not interact with one another, and, on the other hand, that they can be used as a source for applications such as spectroscopy, benefiting from an improved signal-to-noise ratio.

\begin{figure}
\includegraphics[width=8.5cm]{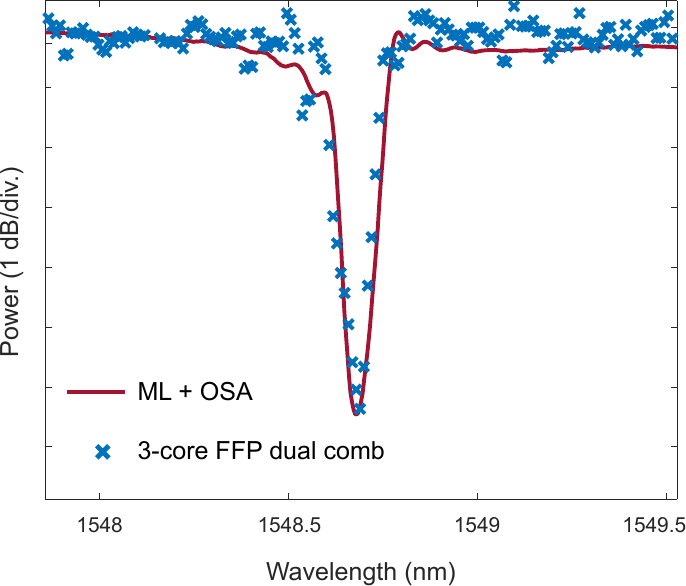}
\caption{Dual-comb spectroscopy proof-of-concept. Measurement of a \textit{Waveshaper} transfer function with mode-locked laser + OSA in red, and with the three-core FFP dual-comb source + photodiode in blue.}
\label{spectro}
\end{figure}

As a proof-of-concept, we record the transfer function of a tunable optical filter (\textit{Waveshaper}) using a symmetric configuration, where the Waveshaper is placed after the recombination of the two OFCs [Fig.~\ref{dc}(a)]. 
A 0.1~nm-wide (12.5~GHz) absorption band (FWHM) is characterized in two different ways: first, with a commercial mode-locked laser source (\textit{Pritel}) and an OSA to obtain a reference curve, and second, with the dual-comb setup. The optical spectrum corresponding to the measured RF spectrum is retrieved using the relation $f_{opt}=m({f_{RF}-f_0})+\frac{c}{\lambda_{pump}}$ where $f_{opt}$ are the optical frequencies, $f_{RF}$ are the RF frequencies, $\lambda_{pump}=1550$~nm is the pump wavelength and $c$ is the speed of light in vacuum. The absorption spectra in Fig.~\ref{spectro} are obtained by normalizing the measured and reference spectra, \textit{i.e.}, by subtracting the measured combs with and without the Waveshaper, and dividing by two for the dual-comb setup due to the symmetric configuration~\cite{coddington_dual-comb_2016}. The dual-comb measurement (blue crosses) shows excellent agreement with the reference curve (red line) in Fig.~\ref{spectro}, highlighting the potential of our dual-comb source implemented in the three-core FFP cavity. While the mode-locked laser + OSA measurement typically requires a few seconds, the dual-comb measurement is completed in less than 1~ms and provides a resolution equal to the cavity FSR (1.27~GHz), \textit{i.e.}, ten times higher than that of a conventional OSA. These results also demonstrate the applicability of multi-soliton regimes, such as the soliton gas of Fig.~\ref{dc}(f), which offer the advantage of higher spectral power. Single-CS regimes could also be exploited for such applications, provided modifications to reach higher comb power. For instance, soliton amplification~\cite{Kuznetsov_Nardi_2025} could be considered, or cavity engineering by enhancing the group-velocity dispersion or reducing the cavity length~\cite{coen_universal_2013}.

\section{Conclusion}
In this study, we report the generation of two OFCs from a single CW pump within a single high-$Q$ fiber Fabry–Perot resonator by exploiting its transverse dimension. We designed and fabricated a three-core FFP resonator that enables the generation of two highly coherent frequency combs with slightly different repetition rates, while the third core is used to lock the pump laser to the cavity. The 30-$\mu$m separation between the cores prevents spurious interactions between the signals (the two combs and the stabilization signal) while ensuring that they experience similar external perturbations, thereby maintaining a high degree of mutual coherence. In addition, the FFP platform offers low insertion loss (less than 4~dB in total) and plug-and-play compatibility with photonic systems thanks to FC/PC connectors. The all-fiber fans provide independent access to each core at both the input and output of the cavity, enabling flexible control of the system parameters. Using this platform, we implemented a proof-of-concept dual-comb experimental setup. Two cavity-soliton-based frequency combs spanning several tens of nanometers were generated simultaneously. By recombining the two OFCs, we successfully retrieved their spectral profiles in the RF domain. Furthermore, we demonstrated that both single-CS and multi-CS OFCs can be generated in the two cores. While the single-CS regime yields smoother spectra and well-defined pulsed interferograms, the multi-soliton regime offers a higher signal-to-noise ratio while maintaining signal stability, enabling spectroscopic measurements of absorption features as narrow as 0.1~nm.

In addition to spectroscopy, this new Kerr resonator architecture could enable other applications, such as ranging and tomographic imaging~\cite{coddington_dual-comb_2016,picque_frequency_2019,coddington_rapid_2009,vicentini_dual-comb_2021}. We have demonstrated that multicore resonators can be effectively used for dual-frequency comb generation. Looking forward, future resonators could be designed to generate three or more OFCs simultaneously, unlocking further potential for advanced metrology~\cite{bancel_all-fiber_2023,lomsadze_tri-comb_2018,lomsadze_frequency_2017,lomsadze_frequency_2017-1}.

This work therefore contributes to the development of new platforms for mutually coherent multi-frequency comb sources in monolithic resonators, with potential applications in fiber-based systems for spectroscopy, ranging, and imaging.

\begin{acknowledgments}
This work was supported by the Agence Nationale de la Recherche (Programme Investissements d’Avenir, FARCO, LABX and VISOPEC projects); European Regional Development Fund (Photonics for Society P4S); Horizon 2020 European Research Council (950618); and the CNRS (IRP LAFONI).
\end{acknowledgments}

\section*{Conflict of interest}
The authors have no conflicts to disclose.

\section*{Data Availability Statement}
The data that support the findings of this study are available from the corresponding author upon reasonable request.

\nocite{*}
\section*{References}
\input{bibli.bbl}

\end{document}

%% file: bibli.bbl
%

%% file: main.bbl
\begin{thebibliography}{77}%
\makeatletter
\providecommand \@ifxundefined [1]{%
 \@ifx{#1\undefined}
}%
\providecommand \@ifnum [1]{%
 \ifnum #1\expandafter \@firstoftwo
 \else \expandafter \@secondoftwo
 \fi
}%
\providecommand \@ifx [1]{%
 \ifx #1\expandafter \@firstoftwo
 \else \expandafter \@secondoftwo
 \fi
}%
\providecommand \natexlab [1]{#1}%
\providecommand \enquote  [1]{``#1''}%
\providecommand \bibnamefont  [1]{#1}%
\providecommand \bibfnamefont [1]{#1}%
\providecommand \citenamefont [1]{#1}%
\providecommand \href@noop [0]{\@secondoftwo}%
\providecommand \href [0]{\begingroup \@sanitize@url \@href}%
\providecommand \@href[1]{\@@startlink{#1}\@@href}%
\providecommand \@@href[1]{\endgroup#1\@@endlink}%
\providecommand \@sanitize@url [0]{\catcode `\\12\catcode `\$12\catcode `\&12\catcode `\#12\catcode `\^12\catcode `\_12\catcode `\%12\relax}%
\providecommand \@@startlink[1]{}%
\providecommand \@@endlink[0]{}%
\providecommand \url  [0]{\begingroup\@sanitize@url \@url }%
\providecommand \@url [1]{\endgroup\@href {#1}{\urlprefix }}%
\providecommand \urlprefix  [0]{URL }%
\providecommand \Eprint [0]{\href }%
\providecommand \doibase [0]{http://dx.doi.org/}%
\providecommand \selectlanguage [0]{\@gobble}%
\providecommand \bibinfo  [0]{\@secondoftwo}%
\providecommand \bibfield  [0]{\@secondoftwo}%
\providecommand \translation [1]{[#1]}%
\providecommand \BibitemOpen [0]{}%
\providecommand \bibitemStop [0]{}%
\providecommand \bibitemNoStop [0]{.\EOS\space}%
\providecommand \EOS [0]{\spacefactor3000\relax}%
\providecommand \BibitemShut  [1]{\csname bibitem#1\endcsname}%
\let\auto@bib@innerbib\@empty
\bibitem [{\citenamefont {Jones}\ \emph {et~al.}(2000)\citenamefont {Jones}, \citenamefont {Diddams}, \citenamefont {Ranka}, \citenamefont {Stentz}, \citenamefont {Windeler}, \citenamefont {Hall},\ and\ \citenamefont {Cundiff}}]{jones_carrier-envelope_2000}%
  \BibitemOpen
  \bibfield  {author} {\bibinfo {author} {\bibfnamefont {D.~J.}\ \bibnamefont {Jones}}, \bibinfo {author} {\bibfnamefont {S.~A.}\ \bibnamefont {Diddams}}, \bibinfo {author} {\bibfnamefont {J.~K.}\ \bibnamefont {Ranka}}, \bibinfo {author} {\bibfnamefont {A.}~\bibnamefont {Stentz}}, \bibinfo {author} {\bibfnamefont {R.~S.}\ \bibnamefont {Windeler}}, \bibinfo {author} {\bibfnamefont {J.~L.}\ \bibnamefont {Hall}}, \ and\ \bibinfo {author} {\bibfnamefont {S.~T.}\ \bibnamefont {Cundiff}},\ }\bibfield  {title} {\enquote {\bibinfo {title} {Carrier-{Envelope} {Phase} {Control} of {Femtosecond} {Mode}-{Locked} {Lasers} and {Direct} {Optical} {Frequency} {Synthesis}},}\ }\href {\doibase 10.1126/science.288.5466.635} {\bibfield  {journal} {\bibinfo  {journal} {Science}\ }\textbf {\bibinfo {volume} {288}},\ \bibinfo {pages} {635--639} (\bibinfo {year} {2000})}\BibitemShut {NoStop}%
\bibitem [{\citenamefont {Holzwarth}\ \emph {et~al.}(2000)\citenamefont {Holzwarth}, \citenamefont {Udem}, \citenamefont {Hänsch}, \citenamefont {Knight}, \citenamefont {Wadsworth},\ and\ \citenamefont {Russell}}]{holzwarth_optical_2000}%
  \BibitemOpen
  \bibfield  {author} {\bibinfo {author} {\bibfnamefont {R.}~\bibnamefont {Holzwarth}}, \bibinfo {author} {\bibfnamefont {T.}~\bibnamefont {Udem}}, \bibinfo {author} {\bibfnamefont {T.~W.}\ \bibnamefont {Hänsch}}, \bibinfo {author} {\bibfnamefont {J.~C.}\ \bibnamefont {Knight}}, \bibinfo {author} {\bibfnamefont {W.~J.}\ \bibnamefont {Wadsworth}}, \ and\ \bibinfo {author} {\bibfnamefont {P.~S.~J.}\ \bibnamefont {Russell}},\ }\bibfield  {title} {\enquote {\bibinfo {title} {Optical {Frequency} {Synthesizer} for {Precision} {Spectroscopy}},}\ }\href {\doibase 10.1103/PhysRevLett.85.2264} {\bibfield  {journal} {\bibinfo  {journal} {Physical Review Letters}\ }\textbf {\bibinfo {volume} {85}},\ \bibinfo {pages} {2264--2267} (\bibinfo {year} {2000})}\BibitemShut {NoStop}%
\bibitem [{\citenamefont {Diddams}\ \emph {et~al.}(2001)\citenamefont {Diddams}, \citenamefont {Udem}, \citenamefont {Bergquist}, \citenamefont {Curtis}, \citenamefont {Drullinger}, \citenamefont {Hollberg}, \citenamefont {Itano}, \citenamefont {Lee}, \citenamefont {Oates}, \citenamefont {Vogel},\ and\ \citenamefont {Wineland}}]{diddams_optical_2001}%
  \BibitemOpen
  \bibfield  {author} {\bibinfo {author} {\bibfnamefont {S.~A.}\ \bibnamefont {Diddams}}, \bibinfo {author} {\bibfnamefont {T.}~\bibnamefont {Udem}}, \bibinfo {author} {\bibfnamefont {J.~C.}\ \bibnamefont {Bergquist}}, \bibinfo {author} {\bibfnamefont {E.~A.}\ \bibnamefont {Curtis}}, \bibinfo {author} {\bibfnamefont {R.~E.}\ \bibnamefont {Drullinger}}, \bibinfo {author} {\bibfnamefont {L.}~\bibnamefont {Hollberg}}, \bibinfo {author} {\bibfnamefont {W.~M.}\ \bibnamefont {Itano}}, \bibinfo {author} {\bibfnamefont {W.~D.}\ \bibnamefont {Lee}}, \bibinfo {author} {\bibfnamefont {C.~W.}\ \bibnamefont {Oates}}, \bibinfo {author} {\bibfnamefont {K.~R.}\ \bibnamefont {Vogel}}, \ and\ \bibinfo {author} {\bibfnamefont {D.~J.}\ \bibnamefont {Wineland}},\ }\bibfield  {title} {\enquote {\bibinfo {title} {An {Optical} {Clock} {Based} on a {Single} {Trapped}$^{\textrm{199}}$ {Hg}$^{\textrm{+}}$ {Ion}},}\ }\href {\doibase 10.1126/science.1061171} {\bibfield  {journal} {\bibinfo  {journal} {Science}\ }\textbf {\bibinfo
  {volume} {293}},\ \bibinfo {pages} {825--828} (\bibinfo {year} {2001})}\BibitemShut {NoStop}%
\bibitem [{\citenamefont {Coddington}, \citenamefont {Newbury},\ and\ \citenamefont {Swann}(2016)}]{coddington_dual-comb_2016}%
  \BibitemOpen
  \bibfield  {author} {\bibinfo {author} {\bibfnamefont {I.}~\bibnamefont {Coddington}}, \bibinfo {author} {\bibfnamefont {N.}~\bibnamefont {Newbury}}, \ and\ \bibinfo {author} {\bibfnamefont {W.}~\bibnamefont {Swann}},\ }\bibfield  {title} {\enquote {\bibinfo {title} {Dual-comb spectroscopy},}\ }\href {\doibase 10.1364/OPTICA.3.000414} {\bibfield  {journal} {\bibinfo  {journal} {Optica}\ }\textbf {\bibinfo {volume} {3}},\ \bibinfo {pages} {414} (\bibinfo {year} {2016})}\BibitemShut {NoStop}%
\bibitem [{\citenamefont {Picqué}\ and\ \citenamefont {Hänsch}(2019)}]{picque_frequency_2019}%
  \BibitemOpen
  \bibfield  {author} {\bibinfo {author} {\bibfnamefont {N.}~\bibnamefont {Picqué}}\ and\ \bibinfo {author} {\bibfnamefont {T.~W.}\ \bibnamefont {Hänsch}},\ }\bibfield  {title} {\enquote {\bibinfo {title} {Frequency comb spectroscopy},}\ }\href {\doibase 10.1038/s41566-018-0347-5} {\bibfield  {journal} {\bibinfo  {journal} {Nature Photonics}\ }\textbf {\bibinfo {volume} {13}},\ \bibinfo {pages} {146--157} (\bibinfo {year} {2019})}\BibitemShut {NoStop}%
\bibitem [{\citenamefont {Coddington}\ \emph {et~al.}(2009)\citenamefont {Coddington}, \citenamefont {Swann}, \citenamefont {Nenadovic},\ and\ \citenamefont {Newbury}}]{coddington_rapid_2009}%
  \BibitemOpen
  \bibfield  {author} {\bibinfo {author} {\bibfnamefont {I.}~\bibnamefont {Coddington}}, \bibinfo {author} {\bibfnamefont {W.~C.}\ \bibnamefont {Swann}}, \bibinfo {author} {\bibfnamefont {L.}~\bibnamefont {Nenadovic}}, \ and\ \bibinfo {author} {\bibfnamefont {N.~R.}\ \bibnamefont {Newbury}},\ }\bibfield  {title} {\enquote {\bibinfo {title} {Rapid and precise absolute distance measurements at long range},}\ }\href {\doibase 10.1038/nphoton.2009.94} {\bibfield  {journal} {\bibinfo  {journal} {Nature Photonics}\ }\textbf {\bibinfo {volume} {3}},\ \bibinfo {pages} {351--356} (\bibinfo {year} {2009})}\BibitemShut {NoStop}%
\bibitem [{\citenamefont {Suh}\ and\ \citenamefont {Vahala}(2018)}]{suh_soliton_2018}%
  \BibitemOpen
  \bibfield  {author} {\bibinfo {author} {\bibfnamefont {M.-G.}\ \bibnamefont {Suh}}\ and\ \bibinfo {author} {\bibfnamefont {K.~J.}\ \bibnamefont {Vahala}},\ }\bibfield  {title} {\enquote {\bibinfo {title} {Soliton microcomb range measurement},}\ }\href {\doibase 10.1126/science.aao1968} {\bibfield  {journal} {\bibinfo  {journal} {Science}\ }\textbf {\bibinfo {volume} {359}},\ \bibinfo {pages} {884--887} (\bibinfo {year} {2018})}\BibitemShut {NoStop}%
\bibitem [{\citenamefont {Xu}\ \emph {et~al.}(2025{\natexlab{a}})\citenamefont {Xu}, \citenamefont {Liu}, \citenamefont {Chen}, \citenamefont {Zhang},\ and\ \citenamefont {Zhang}}]{xu_review_2025}%
  \BibitemOpen
  \bibfield  {author} {\bibinfo {author} {\bibfnamefont {L.}~\bibnamefont {Xu}}, \bibinfo {author} {\bibfnamefont {C.}~\bibnamefont {Liu}}, \bibinfo {author} {\bibfnamefont {L.}~\bibnamefont {Chen}}, \bibinfo {author} {\bibfnamefont {C.}~\bibnamefont {Zhang}}, \ and\ \bibinfo {author} {\bibfnamefont {X.}~\bibnamefont {Zhang}},\ }\bibfield  {title} {\enquote {\bibinfo {title} {A review of dual-chirped-comb interferometry for fast long-distance ranging},}\ }\href {\doibase 10.1063/5.0241325} {\bibfield  {journal} {\bibinfo  {journal} {APL Photonics}\ }\textbf {\bibinfo {volume} {10}},\ \bibinfo {pages} {031201} (\bibinfo {year} {2025}{\natexlab{a}})}\BibitemShut {NoStop}%
\bibitem [{\citenamefont {Vicentini}\ \emph {et~al.}(2021)\citenamefont {Vicentini}, \citenamefont {Wang}, \citenamefont {Van~Gasse}, \citenamefont {Hänsch},\ and\ \citenamefont {Picqué}}]{vicentini_dual-comb_2021}%
  \BibitemOpen
  \bibfield  {author} {\bibinfo {author} {\bibfnamefont {E.}~\bibnamefont {Vicentini}}, \bibinfo {author} {\bibfnamefont {Z.}~\bibnamefont {Wang}}, \bibinfo {author} {\bibfnamefont {K.}~\bibnamefont {Van~Gasse}}, \bibinfo {author} {\bibfnamefont {T.~W.}\ \bibnamefont {Hänsch}}, \ and\ \bibinfo {author} {\bibfnamefont {N.}~\bibnamefont {Picqué}},\ }\bibfield  {title} {\enquote {\bibinfo {title} {Dual-comb hyperspectral digital holography},}\ }\href {\doibase 10.1038/s41566-021-00892-x} {\bibfield  {journal} {\bibinfo  {journal} {Nature Photonics}\ }\textbf {\bibinfo {volume} {15}},\ \bibinfo {pages} {890--894} (\bibinfo {year} {2021})}\BibitemShut {NoStop}%
\bibitem [{\citenamefont {Martín-Mateos}, \citenamefont {Khan},\ and\ \citenamefont {Bonilla-Manrique}(2020)}]{martin-mateos_direct_2020}%
  \BibitemOpen
  \bibfield  {author} {\bibinfo {author} {\bibfnamefont {P.}~\bibnamefont {Martín-Mateos}}, \bibinfo {author} {\bibfnamefont {F.~U.}\ \bibnamefont {Khan}}, \ and\ \bibinfo {author} {\bibfnamefont {O.~E.}\ \bibnamefont {Bonilla-Manrique}},\ }\bibfield  {title} {\enquote {\bibinfo {title} {Direct hyperspectral dual-comb imaging},}\ }\href {\doibase 10.1364/OPTICA.382887} {\bibfield  {journal} {\bibinfo  {journal} {Optica}\ }\textbf {\bibinfo {volume} {7}},\ \bibinfo {pages} {199} (\bibinfo {year} {2020})}\BibitemShut {NoStop}%
\bibitem [{\citenamefont {Link}\ \emph {et~al.}(2015)\citenamefont {Link}, \citenamefont {Klenner}, \citenamefont {Mangold}, \citenamefont {Zaugg}, \citenamefont {Golling}, \citenamefont {Tilma},\ and\ \citenamefont {Keller}}]{link_dual-comb_2015}%
  \BibitemOpen
  \bibfield  {author} {\bibinfo {author} {\bibfnamefont {S.~M.}\ \bibnamefont {Link}}, \bibinfo {author} {\bibfnamefont {A.}~\bibnamefont {Klenner}}, \bibinfo {author} {\bibfnamefont {M.}~\bibnamefont {Mangold}}, \bibinfo {author} {\bibfnamefont {C.~A.}\ \bibnamefont {Zaugg}}, \bibinfo {author} {\bibfnamefont {M.}~\bibnamefont {Golling}}, \bibinfo {author} {\bibfnamefont {B.~W.}\ \bibnamefont {Tilma}}, \ and\ \bibinfo {author} {\bibfnamefont {U.}~\bibnamefont {Keller}},\ }\bibfield  {title} {\enquote {\bibinfo {title} {Dual-comb modelocked laser},}\ }\href {\doibase 10.1364/OE.23.005521} {\bibfield  {journal} {\bibinfo  {journal} {Optics Express}\ }\textbf {\bibinfo {volume} {23}},\ \bibinfo {pages} {5521} (\bibinfo {year} {2015})}\BibitemShut {NoStop}%
\bibitem [{\citenamefont {Ideguchi}\ \emph {et~al.}(2016)\citenamefont {Ideguchi}, \citenamefont {Nakamura}, \citenamefont {Kobayashi},\ and\ \citenamefont {Goda}}]{ideguchi_kerr-lens_2016}%
  \BibitemOpen
  \bibfield  {author} {\bibinfo {author} {\bibfnamefont {T.}~\bibnamefont {Ideguchi}}, \bibinfo {author} {\bibfnamefont {T.}~\bibnamefont {Nakamura}}, \bibinfo {author} {\bibfnamefont {Y.}~\bibnamefont {Kobayashi}}, \ and\ \bibinfo {author} {\bibfnamefont {K.}~\bibnamefont {Goda}},\ }\bibfield  {title} {\enquote {\bibinfo {title} {Kerr-lens mode-locked bidirectional dual-comb ring laser for broadband dual-comb spectroscopy},}\ }\href {\doibase 10.1364/OPTICA.3.000748} {\bibfield  {journal} {\bibinfo  {journal} {Optica}\ }\textbf {\bibinfo {volume} {3}},\ \bibinfo {pages} {748} (\bibinfo {year} {2016})}\BibitemShut {NoStop}%
\bibitem [{\citenamefont {Li}\ \emph {et~al.}(2020)\citenamefont {Li}, \citenamefont {Xing}, \citenamefont {Kwon}, \citenamefont {Xie}, \citenamefont {Prakash}, \citenamefont {Kim},\ and\ \citenamefont {Huang}}]{Li:20}%
  \BibitemOpen
  \bibfield  {author} {\bibinfo {author} {\bibfnamefont {B.}~\bibnamefont {Li}}, \bibinfo {author} {\bibfnamefont {J.}~\bibnamefont {Xing}}, \bibinfo {author} {\bibfnamefont {D.}~\bibnamefont {Kwon}}, \bibinfo {author} {\bibfnamefont {Y.}~\bibnamefont {Xie}}, \bibinfo {author} {\bibfnamefont {N.}~\bibnamefont {Prakash}}, \bibinfo {author} {\bibfnamefont {J.}~\bibnamefont {Kim}}, \ and\ \bibinfo {author} {\bibfnamefont {S.-W.}\ \bibnamefont {Huang}},\ }\bibfield  {title} {\enquote {\bibinfo {title} {Bidirectional mode-locked all-normal dispersion fiber laser},}\ }\href {\doibase 10.1364/OPTICA.396304} {\bibfield  {journal} {\bibinfo  {journal} {Optica}\ }\textbf {\bibinfo {volume} {7}},\ \bibinfo {pages} {961--964} (\bibinfo {year} {2020})}\BibitemShut {NoStop}%
\bibitem [{\citenamefont {Mehravar}\ \emph {et~al.}(2016)\citenamefont {Mehravar}, \citenamefont {Norwood}, \citenamefont {Peyghambarian},\ and\ \citenamefont {Kieu}}]{mehravar_real-time_2016}%
  \BibitemOpen
  \bibfield  {author} {\bibinfo {author} {\bibfnamefont {S.}~\bibnamefont {Mehravar}}, \bibinfo {author} {\bibfnamefont {R.~A.}\ \bibnamefont {Norwood}}, \bibinfo {author} {\bibfnamefont {N.}~\bibnamefont {Peyghambarian}}, \ and\ \bibinfo {author} {\bibfnamefont {K.}~\bibnamefont {Kieu}},\ }\bibfield  {title} {\enquote {\bibinfo {title} {Real-time dual-comb spectroscopy with a free-running bidirectionally mode-locked fiber laser},}\ }\href {\doibase 10.1063/1.4953400} {\bibfield  {journal} {\bibinfo  {journal} {Applied Physics Letters}\ }\textbf {\bibinfo {volume} {108}},\ \bibinfo {pages} {231104} (\bibinfo {year} {2016})}\BibitemShut {NoStop}%
\bibitem [{\citenamefont {Bancel}\ \emph {et~al.}(2023)\citenamefont {Bancel}, \citenamefont {Genier}, \citenamefont {Santagata}, \citenamefont {Conforti}, \citenamefont {Kudlinski}, \citenamefont {Bouwmans}, \citenamefont {Vanvcincq}, \citenamefont {Labat}, \citenamefont {Cassez},\ and\ \citenamefont {Mussot}}]{bancel_all-fiber_2023}%
  \BibitemOpen
  \bibfield  {author} {\bibinfo {author} {\bibfnamefont {E.-L.}\ \bibnamefont {Bancel}}, \bibinfo {author} {\bibfnamefont {E.}~\bibnamefont {Genier}}, \bibinfo {author} {\bibfnamefont {R.}~\bibnamefont {Santagata}}, \bibinfo {author} {\bibfnamefont {M.}~\bibnamefont {Conforti}}, \bibinfo {author} {\bibfnamefont {A.}~\bibnamefont {Kudlinski}}, \bibinfo {author} {\bibfnamefont {G.}~\bibnamefont {Bouwmans}}, \bibinfo {author} {\bibfnamefont {O.}~\bibnamefont {Vanvcincq}}, \bibinfo {author} {\bibfnamefont {D.}~\bibnamefont {Labat}}, \bibinfo {author} {\bibfnamefont {A.}~\bibnamefont {Cassez}}, \ and\ \bibinfo {author} {\bibfnamefont {A.}~\bibnamefont {Mussot}},\ }\bibfield  {title} {\enquote {\bibinfo {title} {All-fiber frequency agile triple-frequency comb light source},}\ }\href {\doibase 10.1038/s41467-023-43734-w} {\bibfield  {journal} {\bibinfo  {journal} {Nature Communications}\ }\textbf {\bibinfo {volume} {14}},\ \bibinfo {pages} {7953} (\bibinfo {year} {2023})}\BibitemShut {NoStop}%
\bibitem [{\citenamefont {Parriaux}, \citenamefont {Hammani},\ and\ \citenamefont {Millot}(2020)}]{parriaux_electro-optic_2020}%
  \BibitemOpen
  \bibfield  {author} {\bibinfo {author} {\bibfnamefont {A.}~\bibnamefont {Parriaux}}, \bibinfo {author} {\bibfnamefont {K.}~\bibnamefont {Hammani}}, \ and\ \bibinfo {author} {\bibfnamefont {G.}~\bibnamefont {Millot}},\ }\bibfield  {title} {\enquote {\bibinfo {title} {Electro-optic frequency combs},}\ }\href {\doibase 10.1364/AOP.382052} {\bibfield  {journal} {\bibinfo  {journal} {Advances in Optics and Photonics}\ }\textbf {\bibinfo {volume} {12}},\ \bibinfo {pages} {223} (\bibinfo {year} {2020})}\BibitemShut {NoStop}%
\bibitem [{\citenamefont {Zhang}\ \emph {et~al.}(2023)\citenamefont {Zhang}, \citenamefont {Tan}, \citenamefont {Chen}, \citenamefont {Yu}, \citenamefont {Wang}, \citenamefont {Chang}, \citenamefont {Liang}, \citenamefont {Guo}, \citenamefont {Zhou}, \citenamefont {Xia}, \citenamefont {Gong}, \citenamefont {Wong}, \citenamefont {Rao}, \citenamefont {Xiao},\ and\ \citenamefont {Yao}}]{zhang_soliton_2023}%
  \BibitemOpen
  \bibfield  {author} {\bibinfo {author} {\bibfnamefont {H.}~\bibnamefont {Zhang}}, \bibinfo {author} {\bibfnamefont {T.}~\bibnamefont {Tan}}, \bibinfo {author} {\bibfnamefont {H.-J.}\ \bibnamefont {Chen}}, \bibinfo {author} {\bibfnamefont {Y.}~\bibnamefont {Yu}}, \bibinfo {author} {\bibfnamefont {W.}~\bibnamefont {Wang}}, \bibinfo {author} {\bibfnamefont {B.}~\bibnamefont {Chang}}, \bibinfo {author} {\bibfnamefont {Y.}~\bibnamefont {Liang}}, \bibinfo {author} {\bibfnamefont {Y.}~\bibnamefont {Guo}}, \bibinfo {author} {\bibfnamefont {H.}~\bibnamefont {Zhou}}, \bibinfo {author} {\bibfnamefont {H.}~\bibnamefont {Xia}}, \bibinfo {author} {\bibfnamefont {Q.}~\bibnamefont {Gong}}, \bibinfo {author} {\bibfnamefont {C.~W.}\ \bibnamefont {Wong}}, \bibinfo {author} {\bibfnamefont {Y.}~\bibnamefont {Rao}}, \bibinfo {author} {\bibfnamefont {Y.-F.}\ \bibnamefont {Xiao}}, \ and\ \bibinfo {author} {\bibfnamefont {B.}~\bibnamefont {Yao}},\ }\bibfield  {title} {\enquote {\bibinfo {title} {Soliton {Microcombs} {Multiplexing}
  {Using} {Intracavity}-{Stimulated} {Brillouin} {Lasers}},}\ }\href {\doibase 10.1103/PhysRevLett.130.153802} {\bibfield  {journal} {\bibinfo  {journal} {Physical Review Letters}\ }\textbf {\bibinfo {volume} {130}},\ \bibinfo {pages} {153802} (\bibinfo {year} {2023})}\BibitemShut {NoStop}%
\bibitem [{\citenamefont {Lucas}\ \emph {et~al.}(2018)\citenamefont {Lucas}, \citenamefont {Lihachev}, \citenamefont {Bouchand}, \citenamefont {Pavlov}, \citenamefont {Raja}, \citenamefont {Karpov}, \citenamefont {Gorodetsky},\ and\ \citenamefont {Kippenberg}}]{lucas_spatial_2018}%
  \BibitemOpen
  \bibfield  {author} {\bibinfo {author} {\bibfnamefont {E.}~\bibnamefont {Lucas}}, \bibinfo {author} {\bibfnamefont {G.}~\bibnamefont {Lihachev}}, \bibinfo {author} {\bibfnamefont {R.}~\bibnamefont {Bouchand}}, \bibinfo {author} {\bibfnamefont {N.~G.}\ \bibnamefont {Pavlov}}, \bibinfo {author} {\bibfnamefont {A.~S.}\ \bibnamefont {Raja}}, \bibinfo {author} {\bibfnamefont {M.}~\bibnamefont {Karpov}}, \bibinfo {author} {\bibfnamefont {M.~L.}\ \bibnamefont {Gorodetsky}}, \ and\ \bibinfo {author} {\bibfnamefont {T.~J.}\ \bibnamefont {Kippenberg}},\ }\bibfield  {title} {\enquote {\bibinfo {title} {Spatial multiplexing of soliton microcombs},}\ }\href {\doibase 10.1038/s41566-018-0256-7} {\bibfield  {journal} {\bibinfo  {journal} {Nature Photonics}\ }\textbf {\bibinfo {volume} {12}},\ \bibinfo {pages} {699--705} (\bibinfo {year} {2018})}\BibitemShut {NoStop}%
\bibitem [{\citenamefont {Xu}\ \emph {et~al.}(2021)\citenamefont {Xu}, \citenamefont {Erkintalo}, \citenamefont {Lin}, \citenamefont {Coen}, \citenamefont {Ma},\ and\ \citenamefont {Murdoch}}]{xu_dual-microcomb_2021}%
  \BibitemOpen
  \bibfield  {author} {\bibinfo {author} {\bibfnamefont {Y.}~\bibnamefont {Xu}}, \bibinfo {author} {\bibfnamefont {M.}~\bibnamefont {Erkintalo}}, \bibinfo {author} {\bibfnamefont {Y.}~\bibnamefont {Lin}}, \bibinfo {author} {\bibfnamefont {S.}~\bibnamefont {Coen}}, \bibinfo {author} {\bibfnamefont {H.}~\bibnamefont {Ma}}, \ and\ \bibinfo {author} {\bibfnamefont {S.~G.}\ \bibnamefont {Murdoch}},\ }\bibfield  {title} {\enquote {\bibinfo {title} {Dual-microcomb generation in a synchronously driven waveguide ring resonator},}\ }\href {\doibase 10.1364/OL.443153} {\bibfield  {journal} {\bibinfo  {journal} {Optics Letters}\ }\textbf {\bibinfo {volume} {46}},\ \bibinfo {pages} {6002} (\bibinfo {year} {2021})}\BibitemShut {NoStop}%
\bibitem [{\citenamefont {Bunel}\ \emph {et~al.}(2025{\natexlab{a}})\citenamefont {Bunel}, \citenamefont {Chatterjee}, \citenamefont {Lumeau}, \citenamefont {Moreau}, \citenamefont {Conforti},\ and\ \citenamefont {Mussot}}]{bunel_dual-frequency_2025}%
  \BibitemOpen
  \bibfield  {author} {\bibinfo {author} {\bibfnamefont {T.}~\bibnamefont {Bunel}}, \bibinfo {author} {\bibfnamefont {D.}~\bibnamefont {Chatterjee}}, \bibinfo {author} {\bibfnamefont {J.}~\bibnamefont {Lumeau}}, \bibinfo {author} {\bibfnamefont {A.}~\bibnamefont {Moreau}}, \bibinfo {author} {\bibfnamefont {M.}~\bibnamefont {Conforti}}, \ and\ \bibinfo {author} {\bibfnamefont {A.}~\bibnamefont {Mussot}},\ }\bibfield  {title} {\enquote {\bibinfo {title} {Dual-frequency comb in fiber {Fabry}–{Perot} resonator},}\ }\href {\doibase 10.1063/5.0248948} {\bibfield  {journal} {\bibinfo  {journal} {APL Photonics}\ }\textbf {\bibinfo {volume} {10}},\ \bibinfo {pages} {036115} (\bibinfo {year} {2025}{\natexlab{a}})}\BibitemShut {NoStop}%
\bibitem [{\citenamefont {Sun}\ \emph {et~al.}(2023)\citenamefont {Sun}, \citenamefont {Wu}, \citenamefont {Tan}, \citenamefont {Xu}, \citenamefont {Li}, \citenamefont {Morandotti}, \citenamefont {Mitchell},\ and\ \citenamefont {Moss}}]{sun_applications_2023}%
  \BibitemOpen
  \bibfield  {author} {\bibinfo {author} {\bibfnamefont {Y.}~\bibnamefont {Sun}}, \bibinfo {author} {\bibfnamefont {J.}~\bibnamefont {Wu}}, \bibinfo {author} {\bibfnamefont {M.}~\bibnamefont {Tan}}, \bibinfo {author} {\bibfnamefont {X.}~\bibnamefont {Xu}}, \bibinfo {author} {\bibfnamefont {Y.}~\bibnamefont {Li}}, \bibinfo {author} {\bibfnamefont {R.}~\bibnamefont {Morandotti}}, \bibinfo {author} {\bibfnamefont {A.}~\bibnamefont {Mitchell}}, \ and\ \bibinfo {author} {\bibfnamefont {D.~J.}\ \bibnamefont {Moss}},\ }\bibfield  {title} {\enquote {\bibinfo {title} {Applications of optical microcombs},}\ }\href {\doibase 10.1364/AOP.470264} {\bibfield  {journal} {\bibinfo  {journal} {Advances in Optics and Photonics}\ }\textbf {\bibinfo {volume} {15}},\ \bibinfo {pages} {86} (\bibinfo {year} {2023})}\BibitemShut {NoStop}%
\bibitem [{\citenamefont {Pasquazi}\ \emph {et~al.}(2018)\citenamefont {Pasquazi}, \citenamefont {Peccianti}, \citenamefont {Razzari}, \citenamefont {Moss}, \citenamefont {Coen}, \citenamefont {Erkintalo}, \citenamefont {Chembo}, \citenamefont {Hansson}, \citenamefont {Wabnitz}, \citenamefont {Del’Haye}, \citenamefont {Xue}, \citenamefont {Weiner},\ and\ \citenamefont {Morandotti}}]{pasquazi_micro-combs_2018}%
  \BibitemOpen
  \bibfield  {author} {\bibinfo {author} {\bibfnamefont {A.}~\bibnamefont {Pasquazi}}, \bibinfo {author} {\bibfnamefont {M.}~\bibnamefont {Peccianti}}, \bibinfo {author} {\bibfnamefont {L.}~\bibnamefont {Razzari}}, \bibinfo {author} {\bibfnamefont {D.~J.}\ \bibnamefont {Moss}}, \bibinfo {author} {\bibfnamefont {S.}~\bibnamefont {Coen}}, \bibinfo {author} {\bibfnamefont {M.}~\bibnamefont {Erkintalo}}, \bibinfo {author} {\bibfnamefont {Y.~K.}\ \bibnamefont {Chembo}}, \bibinfo {author} {\bibfnamefont {T.}~\bibnamefont {Hansson}}, \bibinfo {author} {\bibfnamefont {S.}~\bibnamefont {Wabnitz}}, \bibinfo {author} {\bibfnamefont {P.}~\bibnamefont {Del’Haye}}, \bibinfo {author} {\bibfnamefont {X.}~\bibnamefont {Xue}}, \bibinfo {author} {\bibfnamefont {A.~M.}\ \bibnamefont {Weiner}}, \ and\ \bibinfo {author} {\bibfnamefont {R.}~\bibnamefont {Morandotti}},\ }\bibfield  {title} {\enquote {\bibinfo {title} {Micro-combs: {A} novel generation of optical sources},}\ }\href {\doibase 10.1016/j.physrep.2017.08.004}
  {\bibfield  {journal} {\bibinfo  {journal} {Physics Reports}\ }\textbf {\bibinfo {volume} {729}},\ \bibinfo {pages} {1--81} (\bibinfo {year} {2018})}\BibitemShut {NoStop}%
\bibitem [{\citenamefont {Lei}\ \emph {et~al.}(2022)\citenamefont {Lei}, \citenamefont {Ye}, \citenamefont {Helgason}, \citenamefont {Fülöp}, \citenamefont {Girardi},\ and\ \citenamefont {Torres-Company}}]{lei_optical_2022}%
  \BibitemOpen
  \bibfield  {author} {\bibinfo {author} {\bibfnamefont {F.}~\bibnamefont {Lei}}, \bibinfo {author} {\bibfnamefont {Z.}~\bibnamefont {Ye}}, \bibinfo {author} {\bibfnamefont {O.~B.}\ \bibnamefont {Helgason}}, \bibinfo {author} {\bibfnamefont {A.}~\bibnamefont {Fülöp}}, \bibinfo {author} {\bibfnamefont {M.}~\bibnamefont {Girardi}}, \ and\ \bibinfo {author} {\bibfnamefont {V.}~\bibnamefont {Torres-Company}},\ }\bibfield  {title} {\enquote {\bibinfo {title} {Optical linewidth of soliton microcombs},}\ }\href {\doibase 10.1038/s41467-022-30726-5} {\bibfield  {journal} {\bibinfo  {journal} {Nature Communications}\ }\textbf {\bibinfo {volume} {13}},\ \bibinfo {pages} {3161} (\bibinfo {year} {2022})}\BibitemShut {NoStop}%
\bibitem [{\citenamefont {Bao}, \citenamefont {Suh},\ and\ \citenamefont {Vahala}(2019)}]{bao_microresonator_2019}%
  \BibitemOpen
  \bibfield  {author} {\bibinfo {author} {\bibfnamefont {C.}~\bibnamefont {Bao}}, \bibinfo {author} {\bibfnamefont {M.-G.}\ \bibnamefont {Suh}}, \ and\ \bibinfo {author} {\bibfnamefont {K.}~\bibnamefont {Vahala}},\ }\bibfield  {title} {\enquote {\bibinfo {title} {Microresonator soliton dual-comb imaging},}\ }\href {\doibase 10.1364/OPTICA.6.001110} {\bibfield  {journal} {\bibinfo  {journal} {Optica}\ }\textbf {\bibinfo {volume} {6}},\ \bibinfo {pages} {1110} (\bibinfo {year} {2019})}\BibitemShut {NoStop}%
\bibitem [{\citenamefont {Dutt}\ \emph {et~al.}(2018)\citenamefont {Dutt}, \citenamefont {Joshi}, \citenamefont {Ji}, \citenamefont {Cardenas}, \citenamefont {Okawachi}, \citenamefont {Luke}, \citenamefont {Gaeta},\ and\ \citenamefont {Lipson}}]{dutt_-chip_2018}%
  \BibitemOpen
  \bibfield  {author} {\bibinfo {author} {\bibfnamefont {A.}~\bibnamefont {Dutt}}, \bibinfo {author} {\bibfnamefont {C.}~\bibnamefont {Joshi}}, \bibinfo {author} {\bibfnamefont {X.}~\bibnamefont {Ji}}, \bibinfo {author} {\bibfnamefont {J.}~\bibnamefont {Cardenas}}, \bibinfo {author} {\bibfnamefont {Y.}~\bibnamefont {Okawachi}}, \bibinfo {author} {\bibfnamefont {K.}~\bibnamefont {Luke}}, \bibinfo {author} {\bibfnamefont {A.~L.}\ \bibnamefont {Gaeta}}, \ and\ \bibinfo {author} {\bibfnamefont {M.}~\bibnamefont {Lipson}},\ }\bibfield  {title} {\enquote {\bibinfo {title} {On-chip dual-comb source for spectroscopy},}\ }\href {\doibase 10.1126/sciadv.1701858} {\bibfield  {journal} {\bibinfo  {journal} {Science Advances}\ }\textbf {\bibinfo {volume} {4}},\ \bibinfo {pages} {e1701858} (\bibinfo {year} {2018})}\BibitemShut {NoStop}%
\bibitem [{\citenamefont {Suh}\ \emph {et~al.}(2016)\citenamefont {Suh}, \citenamefont {Yang}, \citenamefont {Yang}, \citenamefont {Yi},\ and\ \citenamefont {Vahala}}]{suh_microresonator_2016}%
  \BibitemOpen
  \bibfield  {author} {\bibinfo {author} {\bibfnamefont {M.-G.}\ \bibnamefont {Suh}}, \bibinfo {author} {\bibfnamefont {Q.-F.}\ \bibnamefont {Yang}}, \bibinfo {author} {\bibfnamefont {K.~Y.}\ \bibnamefont {Yang}}, \bibinfo {author} {\bibfnamefont {X.}~\bibnamefont {Yi}}, \ and\ \bibinfo {author} {\bibfnamefont {K.~J.}\ \bibnamefont {Vahala}},\ }\bibfield  {title} {\enquote {\bibinfo {title} {Microresonator soliton dual-comb spectroscopy},}\ }\href {\doibase 10.1126/science.aah6516} {\bibfield  {journal} {\bibinfo  {journal} {Science}\ }\textbf {\bibinfo {volume} {354}},\ \bibinfo {pages} {600--603} (\bibinfo {year} {2016})}\BibitemShut {NoStop}%
\bibitem [{\citenamefont {Trocha}\ \emph {et~al.}(2018)\citenamefont {Trocha}, \citenamefont {Karpov}, \citenamefont {Ganin}, \citenamefont {Pfeiffer}, \citenamefont {Kordts}, \citenamefont {Wolf}, \citenamefont {Krockenberger}, \citenamefont {Marin-Palomo}, \citenamefont {Weimann}, \citenamefont {Randel}, \citenamefont {Freude}, \citenamefont {Kippenberg},\ and\ \citenamefont {Koos}}]{trocha_ultrafast_2018}%
  \BibitemOpen
  \bibfield  {author} {\bibinfo {author} {\bibfnamefont {P.}~\bibnamefont {Trocha}}, \bibinfo {author} {\bibfnamefont {M.}~\bibnamefont {Karpov}}, \bibinfo {author} {\bibfnamefont {D.}~\bibnamefont {Ganin}}, \bibinfo {author} {\bibfnamefont {M.~H.~P.}\ \bibnamefont {Pfeiffer}}, \bibinfo {author} {\bibfnamefont {A.}~\bibnamefont {Kordts}}, \bibinfo {author} {\bibfnamefont {S.}~\bibnamefont {Wolf}}, \bibinfo {author} {\bibfnamefont {J.}~\bibnamefont {Krockenberger}}, \bibinfo {author} {\bibfnamefont {P.}~\bibnamefont {Marin-Palomo}}, \bibinfo {author} {\bibfnamefont {C.}~\bibnamefont {Weimann}}, \bibinfo {author} {\bibfnamefont {S.}~\bibnamefont {Randel}}, \bibinfo {author} {\bibfnamefont {W.}~\bibnamefont {Freude}}, \bibinfo {author} {\bibfnamefont {T.~J.}\ \bibnamefont {Kippenberg}}, \ and\ \bibinfo {author} {\bibfnamefont {C.}~\bibnamefont {Koos}},\ }\bibfield  {title} {\enquote {\bibinfo {title} {Ultrafast optical ranging using microresonator soliton frequency combs},}\ }\href {\doibase
  10.1126/science.aao3924} {\bibfield  {journal} {\bibinfo  {journal} {Science}\ }\textbf {\bibinfo {volume} {359}},\ \bibinfo {pages} {887--891} (\bibinfo {year} {2018})}\BibitemShut {NoStop}%
\bibitem [{\citenamefont {Rebolledo-Salgado}\ \emph {et~al.}(2023)\citenamefont {Rebolledo-Salgado}, \citenamefont {Quevedo-Galán}, \citenamefont {Helgason}, \citenamefont {Lööf}, \citenamefont {Ye}, \citenamefont {Lei}, \citenamefont {Schröder}, \citenamefont {Zelan},\ and\ \citenamefont {Torres-Company}}]{Rebolledo_2023}%
  \BibitemOpen
  \bibfield  {author} {\bibinfo {author} {\bibfnamefont {I.}~\bibnamefont {Rebolledo-Salgado}}, \bibinfo {author} {\bibfnamefont {C.}~\bibnamefont {Quevedo-Galán}}, \bibinfo {author} {\bibfnamefont {O.~B.}\ \bibnamefont {Helgason}}, \bibinfo {author} {\bibfnamefont {A.}~\bibnamefont {Lööf}}, \bibinfo {author} {\bibfnamefont {Z.}~\bibnamefont {Ye}}, \bibinfo {author} {\bibfnamefont {F.}~\bibnamefont {Lei}}, \bibinfo {author} {\bibfnamefont {J.}~\bibnamefont {Schröder}}, \bibinfo {author} {\bibfnamefont {M.}~\bibnamefont {Zelan}}, \ and\ \bibinfo {author} {\bibfnamefont {V.}~\bibnamefont {Torres-Company}},\ }\bibfield  {title} {\enquote {\bibinfo {title} {Platicon dynamics in photonic molecules},}\ }\href {\doibase 10.1038/s42005-023-01424-5} {\bibfield  {journal} {\bibinfo  {journal} {Communications Physics}\ }\textbf {\bibinfo {volume} {6}} (\bibinfo {year} {2023}),\ 10.1038/s42005-023-01424-5}\BibitemShut {NoStop}%
\bibitem [{\citenamefont {Bourcier}\ \emph {et~al.}(2024)\citenamefont {Bourcier}, \citenamefont {Ousaid}, \citenamefont {Balac}, \citenamefont {Lumeau}, \citenamefont {Moreau}, \citenamefont {Bunel}, \citenamefont {Mussot}, \citenamefont {Conforti}, \citenamefont {Llopis},\ and\ \citenamefont {Fernandez}}]{bourcier_optimization_2024}%
  \BibitemOpen
  \bibfield  {author} {\bibinfo {author} {\bibfnamefont {G.}~\bibnamefont {Bourcier}}, \bibinfo {author} {\bibfnamefont {S.~M.}\ \bibnamefont {Ousaid}}, \bibinfo {author} {\bibfnamefont {S.}~\bibnamefont {Balac}}, \bibinfo {author} {\bibfnamefont {J.}~\bibnamefont {Lumeau}}, \bibinfo {author} {\bibfnamefont {A.}~\bibnamefont {Moreau}}, \bibinfo {author} {\bibfnamefont {T.}~\bibnamefont {Bunel}}, \bibinfo {author} {\bibfnamefont {A.}~\bibnamefont {Mussot}}, \bibinfo {author} {\bibfnamefont {M.}~\bibnamefont {Conforti}}, \bibinfo {author} {\bibfnamefont {O.}~\bibnamefont {Llopis}}, \ and\ \bibinfo {author} {\bibfnamefont {A.}~\bibnamefont {Fernandez}},\ }\bibfield  {title} {\enquote {\bibinfo {title} {Optimization of a fiber {Fabry}–{Perot} resonator for low-threshold modulation instability {Kerr} frequency combs},}\ }\href {\doibase 10.1364/OL.523291} {\bibfield  {journal} {\bibinfo  {journal} {Optics Letters}\ }\textbf {\bibinfo {volume} {49}},\ \bibinfo {pages} {3214} (\bibinfo {year} {2024})}\BibitemShut
  {NoStop}%
\bibitem [{\citenamefont {Bourcier}\ \emph {et~al.}(2025)\citenamefont {Bourcier}, \citenamefont {Balac}, \citenamefont {Mohand-Ousaid}, \citenamefont {Lumeau}, \citenamefont {Moreau}, \citenamefont {Crozatier}, \citenamefont {Sillard}, \citenamefont {Bigot}, \citenamefont {Bigot}, \citenamefont {Llopis},\ and\ \citenamefont {Fernandez}}]{bourcier_investigation_2025}%
  \BibitemOpen
  \bibfield  {author} {\bibinfo {author} {\bibfnamefont {G.}~\bibnamefont {Bourcier}}, \bibinfo {author} {\bibfnamefont {S.}~\bibnamefont {Balac}}, \bibinfo {author} {\bibfnamefont {S.}~\bibnamefont {Mohand-Ousaid}}, \bibinfo {author} {\bibfnamefont {J.}~\bibnamefont {Lumeau}}, \bibinfo {author} {\bibfnamefont {A.}~\bibnamefont {Moreau}}, \bibinfo {author} {\bibfnamefont {V.}~\bibnamefont {Crozatier}}, \bibinfo {author} {\bibfnamefont {P.}~\bibnamefont {Sillard}}, \bibinfo {author} {\bibfnamefont {M.}~\bibnamefont {Bigot}}, \bibinfo {author} {\bibfnamefont {L.}~\bibnamefont {Bigot}}, \bibinfo {author} {\bibfnamefont {O.}~\bibnamefont {Llopis}}, \ and\ \bibinfo {author} {\bibfnamefont {A.}~\bibnamefont {Fernandez}},\ }\bibfield  {title} {\enquote {\bibinfo {title} {Investigation of intrinsic properties of high-quality fiber {Fabry}–{Perot} resonators},}\ }\href {\doibase 10.1364/OL.570921} {\bibfield  {journal} {\bibinfo  {journal} {Optics Letters}\ }\textbf {\bibinfo {volume} {50}},\ \bibinfo {pages} {5566}
  (\bibinfo {year} {2025})}\BibitemShut {NoStop}%
\bibitem [{\citenamefont {Obrzud}, \citenamefont {Lecomte},\ and\ \citenamefont {Herr}(2017)}]{obrzud_temporal_2017}%
  \BibitemOpen
  \bibfield  {author} {\bibinfo {author} {\bibfnamefont {E.}~\bibnamefont {Obrzud}}, \bibinfo {author} {\bibfnamefont {S.}~\bibnamefont {Lecomte}}, \ and\ \bibinfo {author} {\bibfnamefont {T.}~\bibnamefont {Herr}},\ }\bibfield  {title} {\enquote {\bibinfo {title} {Temporal solitons in microresonators driven by optical pulses},}\ }\href {\doibase 10.1038/nphoton.2017.140} {\bibfield  {journal} {\bibinfo  {journal} {Nature Photonics}\ }\textbf {\bibinfo {volume} {11}},\ \bibinfo {pages} {600--607} (\bibinfo {year} {2017})}\BibitemShut {NoStop}%
\bibitem [{\citenamefont {Nie}\ \emph {et~al.}(2022)\citenamefont {Nie}, \citenamefont {Jia}, \citenamefont {Xie}, \citenamefont {Zhu}, \citenamefont {Xie},\ and\ \citenamefont {Huang}}]{nie_synthesized_2022}%
  \BibitemOpen
  \bibfield  {author} {\bibinfo {author} {\bibfnamefont {M.}~\bibnamefont {Nie}}, \bibinfo {author} {\bibfnamefont {K.}~\bibnamefont {Jia}}, \bibinfo {author} {\bibfnamefont {Y.}~\bibnamefont {Xie}}, \bibinfo {author} {\bibfnamefont {S.}~\bibnamefont {Zhu}}, \bibinfo {author} {\bibfnamefont {Z.}~\bibnamefont {Xie}}, \ and\ \bibinfo {author} {\bibfnamefont {S.-W.}\ \bibnamefont {Huang}},\ }\bibfield  {title} {\enquote {\bibinfo {title} {Synthesized spatiotemporal mode-locking and photonic flywheel in multimode mesoresonators},}\ }\href {\doibase 10.1038/s41467-022-34103-0} {\bibfield  {journal} {\bibinfo  {journal} {Nat Commun}\ }\textbf {\bibinfo {volume} {13}},\ \bibinfo {pages} {6395} (\bibinfo {year} {2022})}\BibitemShut {NoStop}%
\bibitem [{\citenamefont {Li}\ \emph {et~al.}(2023)\citenamefont {Li}, \citenamefont {Xu}, \citenamefont {Shamailov}, \citenamefont {Wen}, \citenamefont {Wang}, \citenamefont {Wei}, \citenamefont {Yang}, \citenamefont {Coen}, \citenamefont {Murdoch},\ and\ \citenamefont {Erkintalo}}]{li_ultrashort_2023}%
  \BibitemOpen
  \bibfield  {author} {\bibinfo {author} {\bibfnamefont {Z.}~\bibnamefont {Li}}, \bibinfo {author} {\bibfnamefont {Y.}~\bibnamefont {Xu}}, \bibinfo {author} {\bibfnamefont {S.}~\bibnamefont {Shamailov}}, \bibinfo {author} {\bibfnamefont {X.}~\bibnamefont {Wen}}, \bibinfo {author} {\bibfnamefont {W.}~\bibnamefont {Wang}}, \bibinfo {author} {\bibfnamefont {X.}~\bibnamefont {Wei}}, \bibinfo {author} {\bibfnamefont {Z.}~\bibnamefont {Yang}}, \bibinfo {author} {\bibfnamefont {S.}~\bibnamefont {Coen}}, \bibinfo {author} {\bibfnamefont {S.~G.}\ \bibnamefont {Murdoch}}, \ and\ \bibinfo {author} {\bibfnamefont {M.}~\bibnamefont {Erkintalo}},\ }\bibfield  {title} {\enquote {\bibinfo {title} {Ultrashort dissipative {Raman} solitons in {Kerr} resonators driven with phase-coherent optical pulses},}\ }\href {\doibase 10.1038/s41566-023-01303-z} {\bibfield  {journal} {\bibinfo  {journal} {Nature Photonics}\ } (\bibinfo {year} {2023}),\ 10.1038/s41566-023-01303-z}\BibitemShut {NoStop}%
\bibitem [{\citenamefont {Bunel}\ \emph {et~al.}(2024)\citenamefont {Bunel}, \citenamefont {Conforti}, \citenamefont {Ziani}, \citenamefont {Lumeau}, \citenamefont {Moreau}, \citenamefont {Fernandez}, \citenamefont {Llopis}, \citenamefont {Bourcier},\ and\ \citenamefont {Mussot}}]{bunel_28_2024}%
  \BibitemOpen
  \bibfield  {author} {\bibinfo {author} {\bibfnamefont {T.}~\bibnamefont {Bunel}}, \bibinfo {author} {\bibfnamefont {M.}~\bibnamefont {Conforti}}, \bibinfo {author} {\bibfnamefont {Z.}~\bibnamefont {Ziani}}, \bibinfo {author} {\bibfnamefont {J.}~\bibnamefont {Lumeau}}, \bibinfo {author} {\bibfnamefont {A.}~\bibnamefont {Moreau}}, \bibinfo {author} {\bibfnamefont {A.}~\bibnamefont {Fernandez}}, \bibinfo {author} {\bibfnamefont {O.}~\bibnamefont {Llopis}}, \bibinfo {author} {\bibfnamefont {G.}~\bibnamefont {Bourcier}}, \ and\ \bibinfo {author} {\bibfnamefont {A.}~\bibnamefont {Mussot}},\ }\bibfield  {title} {\enquote {\bibinfo {title} {28 {THz} soliton frequency comb in a continuous-wave pumped fiber {Fabry}–{Pérot} resonator},}\ }\href {\doibase 10.1063/5.0176533} {\bibfield  {journal} {\bibinfo  {journal} {APL Photonics}\ }\textbf {\bibinfo {volume} {9}},\ \bibinfo {pages} {010804} (\bibinfo {year} {2024})}\BibitemShut {NoStop}%
\bibitem [{\citenamefont {Bunel}\ \emph {et~al.}(2025{\natexlab{b}})\citenamefont {Bunel}, \citenamefont {Lumeau}, \citenamefont {Moreau}, \citenamefont {Fernandez}, \citenamefont {Llopis}, \citenamefont {Bourcier}, \citenamefont {Perego}, \citenamefont {Conforti},\ and\ \citenamefont {Mussot}}]{bunel_brillouin-induced_2025}%
  \BibitemOpen
  \bibfield  {author} {\bibinfo {author} {\bibfnamefont {T.}~\bibnamefont {Bunel}}, \bibinfo {author} {\bibfnamefont {J.}~\bibnamefont {Lumeau}}, \bibinfo {author} {\bibfnamefont {A.}~\bibnamefont {Moreau}}, \bibinfo {author} {\bibfnamefont {A.}~\bibnamefont {Fernandez}}, \bibinfo {author} {\bibfnamefont {O.}~\bibnamefont {Llopis}}, \bibinfo {author} {\bibfnamefont {G.}~\bibnamefont {Bourcier}}, \bibinfo {author} {\bibfnamefont {A.~M.}\ \bibnamefont {Perego}}, \bibinfo {author} {\bibfnamefont {M.}~\bibnamefont {Conforti}}, \ and\ \bibinfo {author} {\bibfnamefont {A.}~\bibnamefont {Mussot}},\ }\bibfield  {title} {\enquote {\bibinfo {title} {Brillouin-induced {Kerr} frequency comb in normal dispersion fiber {Fabry} {Perot} resonators},}\ }\href {\doibase 10.1038/s41467-025-60261-y} {\bibfield  {journal} {\bibinfo  {journal} {Nature Communications}\ }\textbf {\bibinfo {volume} {16}},\ \bibinfo {pages} {5160} (\bibinfo {year} {2025}{\natexlab{b}})}\BibitemShut {NoStop}%
\bibitem [{\citenamefont {Xu}\ \emph {et~al.}(2025{\natexlab{b}})\citenamefont {Xu}, \citenamefont {Coen}, \citenamefont {Erkintalo},\ and\ \citenamefont {Murdoch}}]{xu_toward_2025}%
  \BibitemOpen
  \bibfield  {author} {\bibinfo {author} {\bibfnamefont {Y.}~\bibnamefont {Xu}}, \bibinfo {author} {\bibfnamefont {S.}~\bibnamefont {Coen}}, \bibinfo {author} {\bibfnamefont {M.}~\bibnamefont {Erkintalo}}, \ and\ \bibinfo {author} {\bibfnamefont {S.~G.}\ \bibnamefont {Murdoch}},\ }\bibfield  {title} {\enquote {\bibinfo {title} {Toward visible ultrafast imaging with a synchronously pumped switching wave {Kerr} frequency comb},}\ }\href {\doibase 10.1364/OE.551627} {\bibfield  {journal} {\bibinfo  {journal} {Optics Express}\ }\textbf {\bibinfo {volume} {33}},\ \bibinfo {pages} {4714} (\bibinfo {year} {2025}{\natexlab{b}})}\BibitemShut {NoStop}%
\bibitem [{\citenamefont {Qin}\ \emph {et~al.}(2025)\citenamefont {Qin}, \citenamefont {Jia}, \citenamefont {Zhao}, \citenamefont {Ji}, \citenamefont {Shi}, \citenamefont {Zhang}, \citenamefont {Ji}, \citenamefont {Yi}, \citenamefont {Shi}, \citenamefont {Wang}, \citenamefont {Jiang}, \citenamefont {Jin}, \citenamefont {ning Zhu}, \citenamefont {Liang},\ and\ \citenamefont {Xie}}]{Qin:25}%
  \BibitemOpen
  \bibfield  {author} {\bibinfo {author} {\bibfnamefont {C.}~\bibnamefont {Qin}}, \bibinfo {author} {\bibfnamefont {K.}~\bibnamefont {Jia}}, \bibinfo {author} {\bibfnamefont {Z.}~\bibnamefont {Zhao}}, \bibinfo {author} {\bibfnamefont {Y.}~\bibnamefont {Ji}}, \bibinfo {author} {\bibfnamefont {Y.}~\bibnamefont {Shi}}, \bibinfo {author} {\bibfnamefont {X.}~\bibnamefont {Zhang}}, \bibinfo {author} {\bibfnamefont {J.}~\bibnamefont {Ji}}, \bibinfo {author} {\bibfnamefont {X.}~\bibnamefont {Yi}}, \bibinfo {author} {\bibfnamefont {H.}~\bibnamefont {Shi}}, \bibinfo {author} {\bibfnamefont {K.}~\bibnamefont {Wang}}, \bibinfo {author} {\bibfnamefont {X.}~\bibnamefont {Jiang}}, \bibinfo {author} {\bibfnamefont {B.}~\bibnamefont {Jin}}, \bibinfo {author} {\bibfnamefont {S.}~\bibnamefont {ning Zhu}}, \bibinfo {author} {\bibfnamefont {W.}~\bibnamefont {Liang}}, \ and\ \bibinfo {author} {\bibfnamefont {Z.}~\bibnamefont {Xie}},\ }\bibfield  {title} {\enquote {\bibinfo {title} {Compact low-noise dual microcombs for
  high-precision ranging and spectroscopy applications},}\ }\href {\doibase 10.1364/OPTICA.565936} {\bibfield  {journal} {\bibinfo  {journal} {Optica}\ }\textbf {\bibinfo {volume} {12}},\ \bibinfo {pages} {1747--1756} (\bibinfo {year} {2025})}\BibitemShut {NoStop}%
\bibitem [{\citenamefont {Morioka}\ \emph {et~al.}(2012)\citenamefont {Morioka}, \citenamefont {Awaji}, \citenamefont {Ryf}, \citenamefont {Winzer}, \citenamefont {Richardson},\ and\ \citenamefont {Poletti}}]{morioka_enhancing_2012}%
  \BibitemOpen
  \bibfield  {author} {\bibinfo {author} {\bibfnamefont {T.}~\bibnamefont {Morioka}}, \bibinfo {author} {\bibfnamefont {Y.}~\bibnamefont {Awaji}}, \bibinfo {author} {\bibfnamefont {R.}~\bibnamefont {Ryf}}, \bibinfo {author} {\bibfnamefont {P.}~\bibnamefont {Winzer}}, \bibinfo {author} {\bibfnamefont {D.}~\bibnamefont {Richardson}}, \ and\ \bibinfo {author} {\bibfnamefont {F.}~\bibnamefont {Poletti}},\ }\bibfield  {title} {\enquote {\bibinfo {title} {Enhancing optical communications with brand new fibers},}\ }\href {\doibase 10.1109/MCOM.2012.6146483} {\bibfield  {journal} {\bibinfo  {journal} {IEEE Communications Magazine}\ }\textbf {\bibinfo {volume} {50}},\ \bibinfo {pages} {s31--s42} (\bibinfo {year} {2012})}\BibitemShut {NoStop}%
\bibitem [{\citenamefont {Chatterjee}\ \emph {et~al.}(2025{\natexlab{a}})\citenamefont {Chatterjee}, \citenamefont {Bancel}, \citenamefont {Conforti}, \citenamefont {Sivankutty}, \citenamefont {Rigneault}, \citenamefont {Szriftgiser}, \citenamefont {Cundiff},\ and\ \citenamefont {Mussot}}]{chatterjee_sensitivity_2025}%
  \BibitemOpen
  \bibfield  {author} {\bibinfo {author} {\bibfnamefont {D.}~\bibnamefont {Chatterjee}}, \bibinfo {author} {\bibfnamefont {E.-L.}\ \bibnamefont {Bancel}}, \bibinfo {author} {\bibfnamefont {M.}~\bibnamefont {Conforti}}, \bibinfo {author} {\bibfnamefont {S.}~\bibnamefont {Sivankutty}}, \bibinfo {author} {\bibfnamefont {H.}~\bibnamefont {Rigneault}}, \bibinfo {author} {\bibfnamefont {P.}~\bibnamefont {Szriftgiser}}, \bibinfo {author} {\bibfnamefont {S.}~\bibnamefont {Cundiff}}, \ and\ \bibinfo {author} {\bibfnamefont {A.}~\bibnamefont {Mussot}},\ }\bibfield  {title} {\enquote {\bibinfo {title} {Sensitivity {Improvement} in {Dual} {Comb} {Spectroscopy} {With} {Time}-{Programmed} {Electro}-{Optic} {Frequency} {Combs}},}\ }\href {\doibase 10.1109/JLT.2025.3544461} {\bibfield  {journal} {\bibinfo  {journal} {Journal of Lightwave Technology}\ }\textbf {\bibinfo {volume} {43}},\ \bibinfo {pages} {4648--4658} (\bibinfo {year} {2025}{\natexlab{a}})}\BibitemShut {NoStop}%
\bibitem [{\citenamefont {Chatterjee}\ \emph {et~al.}(2025{\natexlab{b}})\citenamefont {Chatterjee}, \citenamefont {Parriaux}, \citenamefont {boivinet}, \citenamefont {bowmans}, \citenamefont {LABAT}, \citenamefont {Cassez},\ and\ \citenamefont {Mussot}}]{Chatterjee2025}%
  \BibitemOpen
  \bibfield  {author} {\bibinfo {author} {\bibfnamefont {D.}~\bibnamefont {Chatterjee}}, \bibinfo {author} {\bibfnamefont {A.}~\bibnamefont {Parriaux}}, \bibinfo {author} {\bibfnamefont {S.}~\bibnamefont {boivinet}}, \bibinfo {author} {\bibfnamefont {G.}~\bibnamefont {bowmans}}, \bibinfo {author} {\bibfnamefont {D.}~\bibnamefont {LABAT}}, \bibinfo {author} {\bibfnamefont {A.}~\bibnamefont {Cassez}}, \ and\ \bibinfo {author} {\bibfnamefont {A.}~\bibnamefont {Mussot}},\ }\href {\doibase 10.1364/opticaopen.29631305.v1} {\enquote {\bibinfo {title} {{Comparative noise analysis of nonlinearly broadened dual frequency combs in different fiber propagation schemes}},}\ } (\bibinfo {year} {2025}{\natexlab{b}})\BibitemShut {NoStop}%
\bibitem [{\citenamefont {Englebert}\ \emph {et~al.}(2023)\citenamefont {Englebert}, \citenamefont {Arabí}, \citenamefont {Gorza},\ and\ \citenamefont {Leo}}]{englebert_high_2023}%
  \BibitemOpen
  \bibfield  {author} {\bibinfo {author} {\bibfnamefont {N.}~\bibnamefont {Englebert}}, \bibinfo {author} {\bibfnamefont {C.~M.}\ \bibnamefont {Arabí}}, \bibinfo {author} {\bibfnamefont {S.-P.}\ \bibnamefont {Gorza}}, \ and\ \bibinfo {author} {\bibfnamefont {F.}~\bibnamefont {Leo}},\ }\bibfield  {title} {\enquote {\bibinfo {title} {High peak-to-background-ratio solitons in a coherently driven active fiber cavity},}\ }\href {\doibase 10.1063/5.0159693} {\bibfield  {journal} {\bibinfo  {journal} {APL Photonics}\ }\textbf {\bibinfo {volume} {8}},\ \bibinfo {pages} {120802} (\bibinfo {year} {2023})}\BibitemShut {NoStop}%
\bibitem [{\citenamefont {Bunel}\ \emph {et~al.}(2023{\natexlab{a}})\citenamefont {Bunel}, \citenamefont {Ziani}, \citenamefont {Conforti}, \citenamefont {Lumeau}, \citenamefont {Moreau}, \citenamefont {Fernandez}, \citenamefont {Llopis}, \citenamefont {Bourcier}, \citenamefont {Perego},\ and\ \citenamefont {Mussot}}]{bunel_impact_2023}%
  \BibitemOpen
  \bibfield  {author} {\bibinfo {author} {\bibfnamefont {T.}~\bibnamefont {Bunel}}, \bibinfo {author} {\bibfnamefont {Z.}~\bibnamefont {Ziani}}, \bibinfo {author} {\bibfnamefont {M.}~\bibnamefont {Conforti}}, \bibinfo {author} {\bibfnamefont {J.}~\bibnamefont {Lumeau}}, \bibinfo {author} {\bibfnamefont {A.}~\bibnamefont {Moreau}}, \bibinfo {author} {\bibfnamefont {A.}~\bibnamefont {Fernandez}}, \bibinfo {author} {\bibfnamefont {O.}~\bibnamefont {Llopis}}, \bibinfo {author} {\bibfnamefont {G.}~\bibnamefont {Bourcier}}, \bibinfo {author} {\bibfnamefont {A.~M.}\ \bibnamefont {Perego}}, \ and\ \bibinfo {author} {\bibfnamefont {A.}~\bibnamefont {Mussot}},\ }\bibfield  {title} {\enquote {\bibinfo {title} {Impact of pump pulse duration on modulation instability {Kerr} frequency combs in fiber {Fabry}–{Pérot} resonators},}\ }\href {\doibase 10.1364/OL.506100} {\bibfield  {journal} {\bibinfo  {journal} {Optics Letters}\ }\textbf {\bibinfo {volume} {48}},\ \bibinfo {pages} {5955} (\bibinfo {year}
  {2023}{\natexlab{a}})}\BibitemShut {NoStop}%
\bibitem [{\citenamefont {Drever}\ \emph {et~al.}(1983)\citenamefont {Drever}, \citenamefont {Hall}, \citenamefont {Kowalski}, \citenamefont {Hough}, \citenamefont {Ford}, \citenamefont {Munley},\ and\ \citenamefont {Ward}}]{drever_laser_1983}%
  \BibitemOpen
  \bibfield  {author} {\bibinfo {author} {\bibfnamefont {R.~W.~P.}\ \bibnamefont {Drever}}, \bibinfo {author} {\bibfnamefont {J.~L.}\ \bibnamefont {Hall}}, \bibinfo {author} {\bibfnamefont {F.~V.}\ \bibnamefont {Kowalski}}, \bibinfo {author} {\bibfnamefont {J.}~\bibnamefont {Hough}}, \bibinfo {author} {\bibfnamefont {G.~M.}\ \bibnamefont {Ford}}, \bibinfo {author} {\bibfnamefont {A.~J.}\ \bibnamefont {Munley}}, \ and\ \bibinfo {author} {\bibfnamefont {H.}~\bibnamefont {Ward}},\ }\bibfield  {title} {\enquote {\bibinfo {title} {Laser phase and frequency stabilization using an optical resonator},}\ }\href@noop {} {\bibfield  {journal} {\bibinfo  {journal} {Applied Physics B Photophysics and Laser Chemistry}\ }\textbf {\bibinfo {volume} {31}},\ \bibinfo {pages} {97--105} (\bibinfo {year} {1983})}\BibitemShut {NoStop}%
\bibitem [{\citenamefont {Black}(2001)}]{black_introduction_2001}%
  \BibitemOpen
  \bibfield  {author} {\bibinfo {author} {\bibfnamefont {E.~D.}\ \bibnamefont {Black}},\ }\bibfield  {title} {\enquote {\bibinfo {title} {An introduction to {Pound}–{Drever}–{Hall} laser frequency stabilization},}\ }\href@noop {} {\bibfield  {journal} {\bibinfo  {journal} {American Journal of Physics}\ }\textbf {\bibinfo {volume} {69}},\ \bibinfo {pages} {79--87} (\bibinfo {year} {2001})}\BibitemShut {NoStop}%
\bibitem [{\citenamefont {Herr}\ \emph {et~al.}(2014)\citenamefont {Herr}, \citenamefont {Brasch}, \citenamefont {Jost}, \citenamefont {Wang}, \citenamefont {Kondratiev}, \citenamefont {Gorodetsky},\ and\ \citenamefont {Kippenberg}}]{herr_temporal_2014}%
  \BibitemOpen
  \bibfield  {author} {\bibinfo {author} {\bibfnamefont {T.}~\bibnamefont {Herr}}, \bibinfo {author} {\bibfnamefont {V.}~\bibnamefont {Brasch}}, \bibinfo {author} {\bibfnamefont {J.~D.}\ \bibnamefont {Jost}}, \bibinfo {author} {\bibfnamefont {C.~Y.}\ \bibnamefont {Wang}}, \bibinfo {author} {\bibfnamefont {N.~M.}\ \bibnamefont {Kondratiev}}, \bibinfo {author} {\bibfnamefont {M.~L.}\ \bibnamefont {Gorodetsky}}, \ and\ \bibinfo {author} {\bibfnamefont {T.~J.}\ \bibnamefont {Kippenberg}},\ }\bibfield  {title} {\enquote {\bibinfo {title} {Temporal solitons in optical microresonators},}\ }\href {\doibase 10.1038/nphoton.2013.343} {\bibfield  {journal} {\bibinfo  {journal} {Nature Photonics}\ }\textbf {\bibinfo {volume} {8}},\ \bibinfo {pages} {145--152} (\bibinfo {year} {2014})}\BibitemShut {NoStop}%
\bibitem [{\citenamefont {Jang}\ \emph {et~al.}(2015)\citenamefont {Jang}, \citenamefont {Erkintalo}, \citenamefont {Coen},\ and\ \citenamefont {Murdoch}}]{jang_temporal_2015}%
  \BibitemOpen
  \bibfield  {author} {\bibinfo {author} {\bibfnamefont {J.~K.}\ \bibnamefont {Jang}}, \bibinfo {author} {\bibfnamefont {M.}~\bibnamefont {Erkintalo}}, \bibinfo {author} {\bibfnamefont {S.}~\bibnamefont {Coen}}, \ and\ \bibinfo {author} {\bibfnamefont {S.~G.}\ \bibnamefont {Murdoch}},\ }\bibfield  {title} {\enquote {\bibinfo {title} {Temporal tweezing of light through the trapping and manipulation of temporal cavity solitons},}\ }\href {\doibase 10.1038/ncomms8370} {\bibfield  {journal} {\bibinfo  {journal} {Nature Communications}\ }\textbf {\bibinfo {volume} {6}},\ \bibinfo {pages} {7370} (\bibinfo {year} {2015})}\BibitemShut {NoStop}%
\bibitem [{\citenamefont {Jia}\ \emph {et~al.}(2020)\citenamefont {Jia}, \citenamefont {Wang}, \citenamefont {Kwon}, \citenamefont {Wang}, \citenamefont {Tsao}, \citenamefont {Liu}, \citenamefont {Ni}, \citenamefont {Guo}, \citenamefont {Yang}, \citenamefont {Jiang}, \citenamefont {Kim}, \citenamefont {Zhu}, \citenamefont {Xie},\ and\ \citenamefont {Huang}}]{jia_photonic_2020}%
  \BibitemOpen
  \bibfield  {author} {\bibinfo {author} {\bibfnamefont {K.}~\bibnamefont {Jia}}, \bibinfo {author} {\bibfnamefont {X.}~\bibnamefont {Wang}}, \bibinfo {author} {\bibfnamefont {D.}~\bibnamefont {Kwon}}, \bibinfo {author} {\bibfnamefont {J.}~\bibnamefont {Wang}}, \bibinfo {author} {\bibfnamefont {E.}~\bibnamefont {Tsao}}, \bibinfo {author} {\bibfnamefont {H.}~\bibnamefont {Liu}}, \bibinfo {author} {\bibfnamefont {X.}~\bibnamefont {Ni}}, \bibinfo {author} {\bibfnamefont {J.}~\bibnamefont {Guo}}, \bibinfo {author} {\bibfnamefont {M.}~\bibnamefont {Yang}}, \bibinfo {author} {\bibfnamefont {X.}~\bibnamefont {Jiang}}, \bibinfo {author} {\bibfnamefont {J.}~\bibnamefont {Kim}}, \bibinfo {author} {\bibfnamefont {S.-n.}\ \bibnamefont {Zhu}}, \bibinfo {author} {\bibfnamefont {Z.}~\bibnamefont {Xie}}, \ and\ \bibinfo {author} {\bibfnamefont {S.-W.}\ \bibnamefont {Huang}},\ }\bibfield  {title} {\enquote {\bibinfo {title} {Photonic {Flywheel} in a {Monolithic} {Fiber} {Resonator}},}\ }\href {\doibase
  10.1103/PhysRevLett.125.143902} {\bibfield  {journal} {\bibinfo  {journal} {Phys. Rev. Lett.}\ }\textbf {\bibinfo {volume} {125}},\ \bibinfo {pages} {143902} (\bibinfo {year} {2020})}\BibitemShut {NoStop}%
\bibitem [{\citenamefont {Cole}\ \emph {et~al.}(2018)\citenamefont {Cole}, \citenamefont {Gatti}, \citenamefont {Papp}, \citenamefont {Prati},\ and\ \citenamefont {Lugiato}}]{cole_theory_2018}%
  \BibitemOpen
  \bibfield  {author} {\bibinfo {author} {\bibfnamefont {D.~C.}\ \bibnamefont {Cole}}, \bibinfo {author} {\bibfnamefont {A.}~\bibnamefont {Gatti}}, \bibinfo {author} {\bibfnamefont {S.~B.}\ \bibnamefont {Papp}}, \bibinfo {author} {\bibfnamefont {F.}~\bibnamefont {Prati}}, \ and\ \bibinfo {author} {\bibfnamefont {L.}~\bibnamefont {Lugiato}},\ }\bibfield  {title} {\enquote {\bibinfo {title} {Theory of kerr frequency combs in fabry-perot resonators},}\ }\href@noop {} {\bibfield  {journal} {\bibinfo  {journal} {Physical Review A}\ }\textbf {\bibinfo {volume} {98}},\ \bibinfo {pages} {013831} (\bibinfo {year} {2018})}\BibitemShut {NoStop}%
\bibitem [{\citenamefont {Ziani}\ \emph {et~al.}(2024)\citenamefont {Ziani}, \citenamefont {Bunel}, \citenamefont {Perego}, \citenamefont {Mussot},\ and\ \citenamefont {Conforti}}]{ziani_theory_2023}%
  \BibitemOpen
  \bibfield  {author} {\bibinfo {author} {\bibfnamefont {Z.}~\bibnamefont {Ziani}}, \bibinfo {author} {\bibfnamefont {T.}~\bibnamefont {Bunel}}, \bibinfo {author} {\bibfnamefont {A.~M.}\ \bibnamefont {Perego}}, \bibinfo {author} {\bibfnamefont {A.}~\bibnamefont {Mussot}}, \ and\ \bibinfo {author} {\bibfnamefont {M.}~\bibnamefont {Conforti}},\ }\bibfield  {title} {\enquote {\bibinfo {title} {{Theory of modulation instability in Kerr Fabry-Perot resonators beyond the mean-field limit}},}\ }\href {\doibase 10.1103/PhysRevA.109.013507} {\bibfield  {journal} {\bibinfo  {journal} {Physical Review A}\ }\textbf {\bibinfo {volume} {109}},\ \bibinfo {pages} {013507} (\bibinfo {year} {2024})}\BibitemShut {NoStop}%
\bibitem [{\citenamefont {Bunel}, \citenamefont {Conforti},\ and\ \citenamefont {Mussot}(2025)}]{bunel_experimental_2025}%
  \BibitemOpen
  \bibfield  {author} {\bibinfo {author} {\bibfnamefont {T.}~\bibnamefont {Bunel}}, \bibinfo {author} {\bibfnamefont {M.}~\bibnamefont {Conforti}}, \ and\ \bibinfo {author} {\bibfnamefont {A.}~\bibnamefont {Mussot}},\ }\bibfield  {title} {\enquote {\bibinfo {title} {Experimental observation of self-frequency-shifting {Raman} quasisolitons in fiber {Fabry}-{Perot} resonators},}\ }\href {\doibase 10.1103/xpk7-n3sv} {\bibfield  {journal} {\bibinfo  {journal} {Physical Review A}\ }\textbf {\bibinfo {volume} {112}},\ \bibinfo {pages} {033535} (\bibinfo {year} {2025})}\BibitemShut {NoStop}%
\bibitem [{\citenamefont {Karpov}\ \emph {et~al.}(2016)\citenamefont {Karpov}, \citenamefont {Guo}, \citenamefont {Kordts}, \citenamefont {Brasch}, \citenamefont {Pfeiffer}, \citenamefont {Zervas}, \citenamefont {Geiselmann},\ and\ \citenamefont {Kippenberg}}]{karpov_raman_2016}%
  \BibitemOpen
  \bibfield  {author} {\bibinfo {author} {\bibfnamefont {M.}~\bibnamefont {Karpov}}, \bibinfo {author} {\bibfnamefont {H.}~\bibnamefont {Guo}}, \bibinfo {author} {\bibfnamefont {A.}~\bibnamefont {Kordts}}, \bibinfo {author} {\bibfnamefont {V.}~\bibnamefont {Brasch}}, \bibinfo {author} {\bibfnamefont {M.~H.}\ \bibnamefont {Pfeiffer}}, \bibinfo {author} {\bibfnamefont {M.}~\bibnamefont {Zervas}}, \bibinfo {author} {\bibfnamefont {M.}~\bibnamefont {Geiselmann}}, \ and\ \bibinfo {author} {\bibfnamefont {T.~J.}\ \bibnamefont {Kippenberg}},\ }\bibfield  {title} {\enquote {\bibinfo {title} {Raman {Self}-{Frequency} {Shift} of {Dissipative} {Kerr} {Solitons} in an {Optical} {Microresonator}},}\ }\href {\doibase 10.1103/PhysRevLett.116.103902} {\bibfield  {journal} {\bibinfo  {journal} {Physical Review Letters}\ }\textbf {\bibinfo {volume} {116}},\ \bibinfo {pages} {103902} (\bibinfo {year} {2016})}\BibitemShut {NoStop}%
\bibitem [{\citenamefont {Yi}\ \emph {et~al.}(2018)\citenamefont {Yi}, \citenamefont {Yang}, \citenamefont {Yang},\ and\ \citenamefont {Vahala}}]{yi_vahala_2018}%
  \BibitemOpen
  \bibfield  {author} {\bibinfo {author} {\bibfnamefont {X.}~\bibnamefont {Yi}}, \bibinfo {author} {\bibfnamefont {Q.-F.}\ \bibnamefont {Yang}}, \bibinfo {author} {\bibfnamefont {K.~Y.}\ \bibnamefont {Yang}}, \ and\ \bibinfo {author} {\bibfnamefont {K.}~\bibnamefont {Vahala}},\ }\bibfield  {title} {\enquote {\bibinfo {title} {Imaging soliton dynamics in optical microcavities},}\ }\href {\doibase 10.1038/s41467-018-06031-5} {\bibfield  {journal} {\bibinfo  {journal} {Nature Communications}\ }\textbf {\bibinfo {volume} {9}} (\bibinfo {year} {2018}),\ 10.1038/s41467-018-06031-5}\BibitemShut {NoStop}%
\bibitem [{\citenamefont {Kuznetsov}\ \emph {et~al.}(2025)\citenamefont {Kuznetsov}, \citenamefont {Nardi}, \citenamefont {Riemensberger}, \citenamefont {Davydova}, \citenamefont {Churaev}, \citenamefont {Seidler},\ and\ \citenamefont {Kippenberg}}]{Kuznetsov_Nardi_2025}%
  \BibitemOpen
  \bibfield  {author} {\bibinfo {author} {\bibfnamefont {N.}~\bibnamefont {Kuznetsov}}, \bibinfo {author} {\bibfnamefont {A.}~\bibnamefont {Nardi}}, \bibinfo {author} {\bibfnamefont {J.}~\bibnamefont {Riemensberger}}, \bibinfo {author} {\bibfnamefont {A.}~\bibnamefont {Davydova}}, \bibinfo {author} {\bibfnamefont {M.}~\bibnamefont {Churaev}}, \bibinfo {author} {\bibfnamefont {P.}~\bibnamefont {Seidler}}, \ and\ \bibinfo {author} {\bibfnamefont {T.~J.}\ \bibnamefont {Kippenberg}},\ }\bibfield  {title} {\enquote {\bibinfo {title} {An ultra-broadband photonic-chip-based parametric amplifier},}\ }\href {\doibase 10.1038/s41586-025-08666-z} {\bibfield  {journal} {\bibinfo  {journal} {Nature}\ }\textbf {\bibinfo {volume} {639}},\ \bibinfo {pages} {928–934} (\bibinfo {year} {2025})}\BibitemShut {NoStop}%
\bibitem [{\citenamefont {Coen}\ and\ \citenamefont {Erkintalo}(2013)}]{coen_universal_2013}%
  \BibitemOpen
  \bibfield  {author} {\bibinfo {author} {\bibfnamefont {S.}~\bibnamefont {Coen}}\ and\ \bibinfo {author} {\bibfnamefont {M.}~\bibnamefont {Erkintalo}},\ }\bibfield  {title} {\enquote {\bibinfo {title} {Universal scaling laws of {Kerr} frequency combs},}\ }\href {\doibase 10.1364/OL.38.001790} {\bibfield  {journal} {\bibinfo  {journal} {Optics Letters}\ }\textbf {\bibinfo {volume} {38}},\ \bibinfo {pages} {1790} (\bibinfo {year} {2013})}\BibitemShut {NoStop}%
\bibitem [{\citenamefont {Lomsadze}, \citenamefont {Smith},\ and\ \citenamefont {Cundiff}(2018)}]{lomsadze_tri-comb_2018}%
  \BibitemOpen
  \bibfield  {author} {\bibinfo {author} {\bibfnamefont {B.}~\bibnamefont {Lomsadze}}, \bibinfo {author} {\bibfnamefont {B.~C.}\ \bibnamefont {Smith}}, \ and\ \bibinfo {author} {\bibfnamefont {S.~T.}\ \bibnamefont {Cundiff}},\ }\bibfield  {title} {\enquote {\bibinfo {title} {Tri-comb spectroscopy},}\ }\href {\doibase 10.1038/s41566-018-0267-4} {\bibfield  {journal} {\bibinfo  {journal} {Nature Photonics}\ }\textbf {\bibinfo {volume} {12}},\ \bibinfo {pages} {676--680} (\bibinfo {year} {2018})}\BibitemShut {NoStop}%
\bibitem [{\citenamefont {Lomsadze}\ and\ \citenamefont {Cundiff}(2017{\natexlab{a}})}]{lomsadze_frequency_2017}%
  \BibitemOpen
  \bibfield  {author} {\bibinfo {author} {\bibfnamefont {B.}~\bibnamefont {Lomsadze}}\ and\ \bibinfo {author} {\bibfnamefont {S.~T.}\ \bibnamefont {Cundiff}},\ }\bibfield  {title} {\enquote {\bibinfo {title} {Frequency comb-based four-wave-mixing spectroscopy},}\ }\href {\doibase 10.1364/OL.42.002346} {\bibfield  {journal} {\bibinfo  {journal} {Optics Letters}\ }\textbf {\bibinfo {volume} {42}},\ \bibinfo {pages} {2346} (\bibinfo {year} {2017}{\natexlab{a}})}\BibitemShut {NoStop}%
\bibitem [{\citenamefont {Lomsadze}\ and\ \citenamefont {Cundiff}(2017{\natexlab{b}})}]{lomsadze_frequency_2017-1}%
  \BibitemOpen
  \bibfield  {author} {\bibinfo {author} {\bibfnamefont {B.}~\bibnamefont {Lomsadze}}\ and\ \bibinfo {author} {\bibfnamefont {S.~T.}\ \bibnamefont {Cundiff}},\ }\bibfield  {title} {\enquote {\bibinfo {title} {Frequency combs enable rapid and high-resolution multidimensional coherent spectroscopy},}\ }\href {\doibase 10.1126/science.aao1090} {\bibfield  {journal} {\bibinfo  {journal} {Science}\ }\textbf {\bibinfo {volume} {357}},\ \bibinfo {pages} {1389--1391} (\bibinfo {year} {2017}{\natexlab{b}})}\BibitemShut {NoStop}%
\bibitem [{\citenamefont {Zideluns}\ \emph {et~al.}(2021)\citenamefont {Zideluns}, \citenamefont {Lemarchand}, \citenamefont {Arhilger}, \citenamefont {Hagedorn},\ and\ \citenamefont {Lumeau}}]{zideluns_automated_2021}%
  \BibitemOpen
  \bibfield  {author} {\bibinfo {author} {\bibfnamefont {J.}~\bibnamefont {Zideluns}}, \bibinfo {author} {\bibfnamefont {F.}~\bibnamefont {Lemarchand}}, \bibinfo {author} {\bibfnamefont {D.}~\bibnamefont {Arhilger}}, \bibinfo {author} {\bibfnamefont {H.}~\bibnamefont {Hagedorn}}, \ and\ \bibinfo {author} {\bibfnamefont {J.}~\bibnamefont {Lumeau}},\ }\bibfield  {title} {\enquote {\bibinfo {title} {Automated optical monitoring wavelength selection for thin-film filters},}\ }\href {\doibase 10.1364/OE.439033} {\bibfield  {journal} {\bibinfo  {journal} {Optics Express}\ }\textbf {\bibinfo {volume} {29}},\ \bibinfo {pages} {33398} (\bibinfo {year} {2021})}\BibitemShut {NoStop}%
\bibitem [{\citenamefont {Englebert}\ \emph {et~al.}(2021)\citenamefont {Englebert}, \citenamefont {Mas~Arabí}, \citenamefont {Parra-Rivas}, \citenamefont {Gorza},\ and\ \citenamefont {Leo}}]{englebert_temporal_2021}%
  \BibitemOpen
  \bibfield  {author} {\bibinfo {author} {\bibfnamefont {N.}~\bibnamefont {Englebert}}, \bibinfo {author} {\bibfnamefont {C.}~\bibnamefont {Mas~Arabí}}, \bibinfo {author} {\bibfnamefont {P.}~\bibnamefont {Parra-Rivas}}, \bibinfo {author} {\bibfnamefont {S.-P.}\ \bibnamefont {Gorza}}, \ and\ \bibinfo {author} {\bibfnamefont {F.}~\bibnamefont {Leo}},\ }\bibfield  {title} {\enquote {\bibinfo {title} {Temporal solitons in a coherently driven active resonator},}\ }\href {\doibase 10.1038/s41566-021-00807-w} {\bibfield  {journal} {\bibinfo  {journal} {Nature Photonics}\ }\textbf {\bibinfo {volume} {15}},\ \bibinfo {pages} {536--541} (\bibinfo {year} {2021})}\BibitemShut {NoStop}%
\bibitem [{\citenamefont {Cole}\ \emph {et~al.}(2017)\citenamefont {Cole}, \citenamefont {Lamb}, \citenamefont {Del’Haye}, \citenamefont {Diddams},\ and\ \citenamefont {Papp}}]{cole_soliton_2017}%
  \BibitemOpen
  \bibfield  {author} {\bibinfo {author} {\bibfnamefont {D.~C.}\ \bibnamefont {Cole}}, \bibinfo {author} {\bibfnamefont {E.~S.}\ \bibnamefont {Lamb}}, \bibinfo {author} {\bibfnamefont {P.}~\bibnamefont {Del’Haye}}, \bibinfo {author} {\bibfnamefont {S.~A.}\ \bibnamefont {Diddams}}, \ and\ \bibinfo {author} {\bibfnamefont {S.~B.}\ \bibnamefont {Papp}},\ }\bibfield  {title} {\enquote {\bibinfo {title} {Soliton crystals in {Kerr} resonators},}\ }\href {\doibase 10.1038/s41566-017-0009-z} {\bibfield  {journal} {\bibinfo  {journal} {Nature Photonics}\ }\textbf {\bibinfo {volume} {11}},\ \bibinfo {pages} {671--676} (\bibinfo {year} {2017})}\BibitemShut {NoStop}%
\bibitem [{\citenamefont {Riemensberger}\ \emph {et~al.}(2022)\citenamefont {Riemensberger}, \citenamefont {Kuznetsov}, \citenamefont {Liu}, \citenamefont {He}, \citenamefont {Wang},\ and\ \citenamefont {Kippenberg}}]{riemensberger_photonic_2022}%
  \BibitemOpen
  \bibfield  {author} {\bibinfo {author} {\bibfnamefont {J.}~\bibnamefont {Riemensberger}}, \bibinfo {author} {\bibfnamefont {N.}~\bibnamefont {Kuznetsov}}, \bibinfo {author} {\bibfnamefont {J.}~\bibnamefont {Liu}}, \bibinfo {author} {\bibfnamefont {J.}~\bibnamefont {He}}, \bibinfo {author} {\bibfnamefont {R.~N.}\ \bibnamefont {Wang}}, \ and\ \bibinfo {author} {\bibfnamefont {T.~J.}\ \bibnamefont {Kippenberg}},\ }\bibfield  {title} {\enquote {\bibinfo {title} {A photonic integrated continuous-travelling-wave parametric amplifier},}\ }\href {\doibase 10.1038/s41586-022-05329-1} {\bibfield  {journal} {\bibinfo  {journal} {Nature}\ }\textbf {\bibinfo {volume} {612}},\ \bibinfo {pages} {56--61} (\bibinfo {year} {2022})}\BibitemShut {NoStop}%
\bibitem [{\citenamefont {Brasch}\ \emph {et~al.}(2019)\citenamefont {Brasch}, \citenamefont {Obrzud}, \citenamefont {Lecomte},\ and\ \citenamefont {Herr}}]{brasch_nonlinear_2019}%
  \BibitemOpen
  \bibfield  {author} {\bibinfo {author} {\bibfnamefont {V.}~\bibnamefont {Brasch}}, \bibinfo {author} {\bibfnamefont {E.}~\bibnamefont {Obrzud}}, \bibinfo {author} {\bibfnamefont {S.}~\bibnamefont {Lecomte}}, \ and\ \bibinfo {author} {\bibfnamefont {T.}~\bibnamefont {Herr}},\ }\bibfield  {title} {\enquote {\bibinfo {title} {Nonlinear filtering of an optical pulse train using dissipative {Kerr} solitons},}\ }\href {\doibase 10.1364/OPTICA.6.001386} {\bibfield  {journal} {\bibinfo  {journal} {Optica}\ }\textbf {\bibinfo {volume} {6}},\ \bibinfo {pages} {1386} (\bibinfo {year} {2019})}\BibitemShut {NoStop}%
\bibitem [{\citenamefont {Fortier}\ and\ \citenamefont {Baumann}(2019)}]{fortier_20_2019}%
  \BibitemOpen
  \bibfield  {author} {\bibinfo {author} {\bibfnamefont {T.}~\bibnamefont {Fortier}}\ and\ \bibinfo {author} {\bibfnamefont {E.}~\bibnamefont {Baumann}},\ }\bibfield  {title} {\enquote {\bibinfo {title} {20 years of developments in optical frequency comb technology and applications},}\ }\href {\doibase 10.1038/s42005-019-0249-y} {\bibfield  {journal} {\bibinfo  {journal} {Communications Physics}\ }\textbf {\bibinfo {volume} {2}},\ \bibinfo {pages} {153} (\bibinfo {year} {2019})}\BibitemShut {NoStop}%
\bibitem [{\citenamefont {Kippenberg}\ \emph {et~al.}(2018)\citenamefont {Kippenberg}, \citenamefont {Gaeta}, \citenamefont {Lipson},\ and\ \citenamefont {Gorodetsky}}]{kippenberg_dissipative_2018}%
  \BibitemOpen
  \bibfield  {author} {\bibinfo {author} {\bibfnamefont {T.~J.}\ \bibnamefont {Kippenberg}}, \bibinfo {author} {\bibfnamefont {A.~L.}\ \bibnamefont {Gaeta}}, \bibinfo {author} {\bibfnamefont {M.}~\bibnamefont {Lipson}}, \ and\ \bibinfo {author} {\bibfnamefont {M.~L.}\ \bibnamefont {Gorodetsky}},\ }\bibfield  {title} {\enquote {\bibinfo {title} {Dissipative {Kerr} solitons in optical microresonators},}\ }\href {\doibase 10.1126/science.aan8083} {\bibfield  {journal} {\bibinfo  {journal} {Science}\ }\textbf {\bibinfo {volume} {361}},\ \bibinfo {pages} {eaan8083} (\bibinfo {year} {2018})}\BibitemShut {NoStop}%
\bibitem [{\citenamefont {Bunel}\ \emph {et~al.}(2023{\natexlab{b}})\citenamefont {Bunel}, \citenamefont {Conforti}, \citenamefont {Ziani}, \citenamefont {Lumeau}, \citenamefont {Moreau}, \citenamefont {Fernandez}, \citenamefont {Llopis}, \citenamefont {Roul}, \citenamefont {Perego}, \citenamefont {Wong},\ and\ \citenamefont {Mussot}}]{bunel_observation_2023}%
  \BibitemOpen
  \bibfield  {author} {\bibinfo {author} {\bibfnamefont {T.}~\bibnamefont {Bunel}}, \bibinfo {author} {\bibfnamefont {M.}~\bibnamefont {Conforti}}, \bibinfo {author} {\bibfnamefont {Z.}~\bibnamefont {Ziani}}, \bibinfo {author} {\bibfnamefont {J.}~\bibnamefont {Lumeau}}, \bibinfo {author} {\bibfnamefont {A.}~\bibnamefont {Moreau}}, \bibinfo {author} {\bibfnamefont {A.}~\bibnamefont {Fernandez}}, \bibinfo {author} {\bibfnamefont {O.}~\bibnamefont {Llopis}}, \bibinfo {author} {\bibfnamefont {J.}~\bibnamefont {Roul}}, \bibinfo {author} {\bibfnamefont {A.~M.}\ \bibnamefont {Perego}}, \bibinfo {author} {\bibfnamefont {K.~K.~Y.}\ \bibnamefont {Wong}}, \ and\ \bibinfo {author} {\bibfnamefont {A.}~\bibnamefont {Mussot}},\ }\bibfield  {title} {\enquote {\bibinfo {title} {Observation of modulation instability {Kerr} frequency combs in a fiber {Fabry}–{Pérot} resonator},}\ }\href {\doibase 10.1364/OL.479466} {\bibfield  {journal} {\bibinfo  {journal} {Optics Letters}\ }\textbf {\bibinfo {volume} {48}},\ \bibinfo
  {pages} {275} (\bibinfo {year} {2023}{\natexlab{b}})}\BibitemShut {NoStop}%
\bibitem [{\citenamefont {Xiao}\ \emph {et~al.}(2023{\natexlab{a}})\citenamefont {Xiao}, \citenamefont {Li}, \citenamefont {Cai}, \citenamefont {Zhang}, \citenamefont {Huang}, \citenamefont {Li}, \citenamefont {Yao}, \citenamefont {Wu},\ and\ \citenamefont {Chen}}]{xiao_near-zero-dispersion_2023}%
  \BibitemOpen
  \bibfield  {author} {\bibinfo {author} {\bibfnamefont {Z.}~\bibnamefont {Xiao}}, \bibinfo {author} {\bibfnamefont {T.}~\bibnamefont {Li}}, \bibinfo {author} {\bibfnamefont {M.}~\bibnamefont {Cai}}, \bibinfo {author} {\bibfnamefont {H.}~\bibnamefont {Zhang}}, \bibinfo {author} {\bibfnamefont {Y.}~\bibnamefont {Huang}}, \bibinfo {author} {\bibfnamefont {C.}~\bibnamefont {Li}}, \bibinfo {author} {\bibfnamefont {B.}~\bibnamefont {Yao}}, \bibinfo {author} {\bibfnamefont {K.}~\bibnamefont {Wu}}, \ and\ \bibinfo {author} {\bibfnamefont {J.}~\bibnamefont {Chen}},\ }\bibfield  {title} {\enquote {\bibinfo {title} {Near-zero-dispersion soliton and broadband modulational instability {Kerr} microcombs in anomalous dispersion},}\ }\href {\doibase 10.1038/s41377-023-01076-8} {\bibfield  {journal} {\bibinfo  {journal} {Light: Science \& Applications}\ }\textbf {\bibinfo {volume} {12}},\ \bibinfo {pages} {33} (\bibinfo {year} {2023}{\natexlab{a}})}\BibitemShut {NoStop}%
\bibitem [{\citenamefont {Torres-Company}\ and\ \citenamefont {Weiner}(2014)}]{torres-company_optical_2014}%
  \BibitemOpen
  \bibfield  {author} {\bibinfo {author} {\bibfnamefont {V.}~\bibnamefont {Torres-Company}}\ and\ \bibinfo {author} {\bibfnamefont {A.~M.}\ \bibnamefont {Weiner}},\ }\bibfield  {title} {\enquote {\bibinfo {title} {Optical frequency comb technology for ultra-broadband radio-frequency photonics: {Optical} frequency comb technology for {RF} photonics},}\ }\href {\doibase 10.1002/lpor.201300126} {\bibfield  {journal} {\bibinfo  {journal} {Laser \& Photonics Reviews}\ }\textbf {\bibinfo {volume} {8}},\ \bibinfo {pages} {368--393} (\bibinfo {year} {2014})}\BibitemShut {NoStop}%
\bibitem [{\citenamefont {Godey}\ \emph {et~al.}(2014)\citenamefont {Godey}, \citenamefont {Balakireva}, \citenamefont {Coillet},\ and\ \citenamefont {Chembo}}]{godey_stability_2014}%
  \BibitemOpen
  \bibfield  {author} {\bibinfo {author} {\bibfnamefont {C.}~\bibnamefont {Godey}}, \bibinfo {author} {\bibfnamefont {I.~V.}\ \bibnamefont {Balakireva}}, \bibinfo {author} {\bibfnamefont {A.}~\bibnamefont {Coillet}}, \ and\ \bibinfo {author} {\bibfnamefont {Y.~K.}\ \bibnamefont {Chembo}},\ }\bibfield  {title} {\enquote {\bibinfo {title} {Stability analysis of the spatiotemporal {Lugiato}-{Lefever} model for {Kerr} optical frequency combs in the anomalous and normal dispersion regimes},}\ }\href {\doibase 10.1103/PhysRevA.89.063814} {\bibfield  {journal} {\bibinfo  {journal} {Physical Review A}\ }\textbf {\bibinfo {volume} {89}},\ \bibinfo {pages} {063814} (\bibinfo {year} {2014})}\BibitemShut {NoStop}%
\bibitem [{\citenamefont {Xue}, \citenamefont {Qi},\ and\ \citenamefont {Weiner}(2016)}]{xue_normal-dispersion_2016}%
  \BibitemOpen
  \bibfield  {author} {\bibinfo {author} {\bibfnamefont {X.}~\bibnamefont {Xue}}, \bibinfo {author} {\bibfnamefont {M.}~\bibnamefont {Qi}}, \ and\ \bibinfo {author} {\bibfnamefont {A.~M.}\ \bibnamefont {Weiner}},\ }\bibfield  {title} {\enquote {\bibinfo {title} {Normal-dispersion microresonator {Kerr} frequency combs},}\ }\href {\doibase 10.1515/nanoph-2016-0016} {\bibfield  {journal} {\bibinfo  {journal} {Nanophotonics}\ }\textbf {\bibinfo {volume} {5}},\ \bibinfo {pages} {244--262} (\bibinfo {year} {2016})}\BibitemShut {NoStop}%
\bibitem [{\citenamefont {Fülöp}\ \emph {et~al.}(2018)\citenamefont {Fülöp}, \citenamefont {Mazur}, \citenamefont {Lorences-Riesgo}, \citenamefont {Helgason}, \citenamefont {Wang}, \citenamefont {Xuan}, \citenamefont {Leaird}, \citenamefont {Qi}, \citenamefont {Andrekson}, \citenamefont {Weiner},\ and\ \citenamefont {Torres-Company}}]{fulop_high-order_2018}%
  \BibitemOpen
  \bibfield  {author} {\bibinfo {author} {\bibfnamefont {A.}~\bibnamefont {Fülöp}}, \bibinfo {author} {\bibfnamefont {M.}~\bibnamefont {Mazur}}, \bibinfo {author} {\bibfnamefont {A.}~\bibnamefont {Lorences-Riesgo}}, \bibinfo {author} {\bibfnamefont {O.~B.}\ \bibnamefont {Helgason}}, \bibinfo {author} {\bibfnamefont {P.-H.}\ \bibnamefont {Wang}}, \bibinfo {author} {\bibfnamefont {Y.}~\bibnamefont {Xuan}}, \bibinfo {author} {\bibfnamefont {D.~E.}\ \bibnamefont {Leaird}}, \bibinfo {author} {\bibfnamefont {M.}~\bibnamefont {Qi}}, \bibinfo {author} {\bibfnamefont {P.~A.}\ \bibnamefont {Andrekson}}, \bibinfo {author} {\bibfnamefont {A.~M.}\ \bibnamefont {Weiner}}, \ and\ \bibinfo {author} {\bibfnamefont {V.}~\bibnamefont {Torres-Company}},\ }\bibfield  {title} {\enquote {\bibinfo {title} {High-order coherent communications using mode-locked dark-pulse {Kerr} combs from microresonators},}\ }\href {\doibase 10.1038/s41467-018-04046-6} {\bibfield  {journal} {\bibinfo  {journal} {Nature Communications}\ }\textbf
  {\bibinfo {volume} {9}},\ \bibinfo {pages} {1598} (\bibinfo {year} {2018})}\BibitemShut {NoStop}%
\bibitem [{\citenamefont {Xiao}\ \emph {et~al.}(2023{\natexlab{b}})\citenamefont {Xiao}, \citenamefont {Wu}, \citenamefont {Zhang}, \citenamefont {Li}, \citenamefont {Cai}, \citenamefont {Huang},\ and\ \citenamefont {Chen}}]{xiao_modeling_2023}%
  \BibitemOpen
  \bibfield  {author} {\bibinfo {author} {\bibfnamefont {Z.}~\bibnamefont {Xiao}}, \bibinfo {author} {\bibfnamefont {K.}~\bibnamefont {Wu}}, \bibinfo {author} {\bibfnamefont {H.}~\bibnamefont {Zhang}}, \bibinfo {author} {\bibfnamefont {T.}~\bibnamefont {Li}}, \bibinfo {author} {\bibfnamefont {M.}~\bibnamefont {Cai}}, \bibinfo {author} {\bibfnamefont {Y.}~\bibnamefont {Huang}}, \ and\ \bibinfo {author} {\bibfnamefont {J.}~\bibnamefont {Chen}},\ }\bibfield  {title} {\enquote {\bibinfo {title} {Modeling the {Kerr} {Comb} of a {Pulse} {Pumped} {F}-{P} {Microresonator} {With} {Normal} {Dispersion}},}\ }\href {\doibase 10.1109/JLT.2023.3300191} {\bibfield  {journal} {\bibinfo  {journal} {Journal of Lightwave Technology}\ }\textbf {\bibinfo {volume} {41}},\ \bibinfo {pages} {7408--7417} (\bibinfo {year} {2023}{\natexlab{b}})}\BibitemShut {NoStop}%
\bibitem [{\citenamefont {Macnaughtan}\ \emph {et~al.}(2023)\citenamefont {Macnaughtan}, \citenamefont {Erkintalo}, \citenamefont {Coen}, \citenamefont {Murdoch},\ and\ \citenamefont {Xu}}]{macnaughtan_temporal_2023}%
  \BibitemOpen
  \bibfield  {author} {\bibinfo {author} {\bibfnamefont {M.}~\bibnamefont {Macnaughtan}}, \bibinfo {author} {\bibfnamefont {M.}~\bibnamefont {Erkintalo}}, \bibinfo {author} {\bibfnamefont {S.}~\bibnamefont {Coen}}, \bibinfo {author} {\bibfnamefont {S.}~\bibnamefont {Murdoch}}, \ and\ \bibinfo {author} {\bibfnamefont {Y.}~\bibnamefont {Xu}},\ }\bibfield  {title} {\enquote {\bibinfo {title} {Temporal characteristics of stationary switching waves in a normal dispersion pulsed-pump fiber cavity},}\ }\href {\doibase 10.1364/OL.492998} {\bibfield  {journal} {\bibinfo  {journal} {Optics Letters}\ }\textbf {\bibinfo {volume} {48}},\ \bibinfo {pages} {4097} (\bibinfo {year} {2023})}\BibitemShut {NoStop}%
\bibitem [{\citenamefont {Xu}\ \emph {et~al.}(2020)\citenamefont {Xu}, \citenamefont {Lin}, \citenamefont {Nielsen}, \citenamefont {Hendry}, \citenamefont {Coen}, \citenamefont {Erkintalo}, \citenamefont {Ma},\ and\ \citenamefont {Murdoch}}]{xu_harmonic_2020}%
  \BibitemOpen
  \bibfield  {author} {\bibinfo {author} {\bibfnamefont {Y.}~\bibnamefont {Xu}}, \bibinfo {author} {\bibfnamefont {Y.}~\bibnamefont {Lin}}, \bibinfo {author} {\bibfnamefont {A.}~\bibnamefont {Nielsen}}, \bibinfo {author} {\bibfnamefont {I.}~\bibnamefont {Hendry}}, \bibinfo {author} {\bibfnamefont {S.}~\bibnamefont {Coen}}, \bibinfo {author} {\bibfnamefont {M.}~\bibnamefont {Erkintalo}}, \bibinfo {author} {\bibfnamefont {H.}~\bibnamefont {Ma}}, \ and\ \bibinfo {author} {\bibfnamefont {S.~G.}\ \bibnamefont {Murdoch}},\ }\bibfield  {title} {\enquote {\bibinfo {title} {Harmonic and rational harmonic driving of microresonator soliton frequency combs},}\ }\href {\doibase 10.1364/OPTICA.392571} {\bibfield  {journal} {\bibinfo  {journal} {Optica}\ }\textbf {\bibinfo {volume} {7}},\ \bibinfo {pages} {940} (\bibinfo {year} {2020})}\BibitemShut {NoStop}%
\bibitem [{\citenamefont {Ding}\ \emph {et~al.}(2023)\citenamefont {Ding}, \citenamefont {Wang}, \citenamefont {Xiong}, \citenamefont {Chen},\ and\ \citenamefont {Xu}}]{ding_single-short-cavity_2023}%
  \BibitemOpen
  \bibfield  {author} {\bibinfo {author} {\bibfnamefont {Z.}~\bibnamefont {Ding}}, \bibinfo {author} {\bibfnamefont {G.}~\bibnamefont {Wang}}, \bibinfo {author} {\bibfnamefont {Y.}~\bibnamefont {Xiong}}, \bibinfo {author} {\bibfnamefont {Y.}~\bibnamefont {Chen}}, \ and\ \bibinfo {author} {\bibfnamefont {F.}~\bibnamefont {Xu}},\ }\bibfield  {title} {\enquote {\bibinfo {title} {Single-short-cavity dual-comb fiber laser with over 120 {kHz} repetition rate difference based on polarization multiplexing},}\ }\href {\doibase 10.1364/OL.501835} {\bibfield  {journal} {\bibinfo  {journal} {Optics Letters}\ }\textbf {\bibinfo {volume} {48}},\ \bibinfo {pages} {5233} (\bibinfo {year} {2023})}\BibitemShut {NoStop}%
\bibitem [{\citenamefont {Bunel}(2024)}]{bunel_broadband_2024}%
  \BibitemOpen
  \bibfield  {author} {\bibinfo {author} {\bibfnamefont {T.}~\bibnamefont {Bunel}},\ }\bibfield  {title} {\enquote {\bibinfo {title} {Broadband {Kerr} frequency comb in fiber {Fabry}-{Perot} resonators induced by switching waves},}\ }\href@noop {} {\bibfield  {journal} {\bibinfo  {journal} {PHYSICAL REVIEW A}\ } (\bibinfo {year} {2024})}\BibitemShut {NoStop}%
\bibitem [{\citenamefont {Mandon}, \citenamefont {Guelachvili},\ and\ \citenamefont {Picqué}(2009)}]{mandon_fourier_2009}%
  \BibitemOpen
  \bibfield  {author} {\bibinfo {author} {\bibfnamefont {J.}~\bibnamefont {Mandon}}, \bibinfo {author} {\bibfnamefont {G.}~\bibnamefont {Guelachvili}}, \ and\ \bibinfo {author} {\bibfnamefont {N.}~\bibnamefont {Picqué}},\ }\bibfield  {title} {\enquote {\bibinfo {title} {Fourier transform spectroscopy with a laser frequency comb},}\ }\href {\doibase 10.1038/nphoton.2008.293} {\bibfield  {journal} {\bibinfo  {journal} {Nature Photonics}\ }\textbf {\bibinfo {volume} {3}},\ \bibinfo {pages} {99--102} (\bibinfo {year} {2009})}\BibitemShut {NoStop}%
\bibitem [{\citenamefont {Diddams}, \citenamefont {Hollberg},\ and\ \citenamefont {Mbele}(2007)}]{diddams_molecular_2007}%
  \BibitemOpen
  \bibfield  {author} {\bibinfo {author} {\bibfnamefont {S.~A.}\ \bibnamefont {Diddams}}, \bibinfo {author} {\bibfnamefont {L.}~\bibnamefont {Hollberg}}, \ and\ \bibinfo {author} {\bibfnamefont {V.}~\bibnamefont {Mbele}},\ }\bibfield  {title} {\enquote {\bibinfo {title} {Molecular fingerprinting with the resolved modes of a femtosecond laser frequency comb},}\ }\href {\doibase 10.1038/nature05524} {\bibfield  {journal} {\bibinfo  {journal} {Nature}\ }\textbf {\bibinfo {volume} {445}},\ \bibinfo {pages} {627--630} (\bibinfo {year} {2007})}\BibitemShut {NoStop}%
\end{thebibliography}
